\documentclass[12pt,a4paper abstracton]{scrartcl}

\usepackage{amssymb,amsmath,amsfonts,mathrsfs}

\usepackage[amsmath, thmmarks, hyperref]{ntheorem}

\usepackage{mathrsfs}

\usepackage{fullpage}

\usepackage{tikz}

\usepackage{color}

\usepackage{microtype}

\usepackage{graphicx}

\usepackage{slashed}

\usepackage[ colorlinks=true, citecolor=blue]{hyperref}

\usepackage{caption}

\theoremstyle{plain}

\theoremheaderfont{\normalfont\sffamily\bfseries}

 \theorembodyfont{\itshape}

 \theoremseparator{:}

\newtheorem{thm}{Theorem}

\usepackage{mathptmx}

\DeclareRobustCommand\hbar{\mathchar'26\mkern-9mu h}

\def\ben{\begin{equation}}

\def\een{\end{equation}}

\def\bena{\begin{eqnarray}}

\def\eena{\end{eqnarray}}

\newtheorem{lemma}{Lemma}

\newtheorem{prop}{Proposition}

\newtheorem{cor}{Corollary}

\newtheorem{definition}{Definition}

\newtheorem{remark}{Remark}

\makeatletter

\newcommand{\openbox}{\leavevmode
  \hbox to.77778em{%
  \hfil\vrule \vbox to.675em{\hrule width.6em\vfil\hrule}%
  \vrule\hfil}}
\gdef\proofSymbol{\openbox}

\newcommand{\proofname}{Proof.}

\newcounter{proof}%

  {\@endtheorem}

\makeatother

\newcommand{\1}{1\!\!\!1}

\newcommand{\al}{\alpha}

\newcommand{\be}{\beta}

\newcommand{\de}{\delta}

\newcommand{\De}{\Delta}

\newcommand{\vep}{\varepsilon}

\newcommand{\ka}{\kappa}

\newcommand{\la}{\lambda}

\newcommand{\si}{\sigma}

\newcommand{\vp}{\varphi}

\newcommand{\alo}{\alpha_0}

\newcommand{\cC}{{\cal C}}

\newcommand{\cH}{{\cal H}}

\newcommand{\cS}{{\cal S}}

\newcommand{\pa}{\partial}

\newcommand{\ti}[1]{\tilde{#1}}

\newcommand{\eq}{\begin{equation}}

\newcommand{\eqe}{\end{equation}}

\newcounter{saveeqn}

\newcommand{\mr}{{\mathbb R}}

\newcommand{\mn}{{\mathbb N}}

\newcommand{\ms}{{\mathbb S}}

\newcommand{\mk}{{\mathbb K}}

\newcommand{\hms}{{{\hat {\mathbb S}}}}

\newcommand{\C}{{\mathcal C}}

\newcommand{\e}{\operatorname{e}}

\renewcommand{\L}{{\mathcal L}}

\renewcommand{\S}{\mathscr{S}}

\renewcommand{\d}{{\textrm d}}

\makeatletter

\newcommand*{\rom}[1]{\expandafter\@slowromancap\romannumeral #1@}

\makeatother

\title{Asymptotically free solutions of the scalar flow equations}

\author{Christoph Kopper\\
{\normalsize {\textit{Centre de Physique Th\'eorique CPHT, CNRS, UMR 7644}}}\\
{\normalsize{\textit{Ecole polytechnique, 
Institut Polytechnique de Paris, 91128 Palaiseau, France}}}}

\begin{document}

\maketitle
\begin{abstract}

\centerline{\textbf{Abstract}}

We analyse solutions of the flow equations of the
renormalisation group for the $\,n$-point Schwinger functions
of scalar field theory in four-dimensional Euclidean
space with nonpolynomial boundary conditions.
We show that there exist UV  asymptotically free
solutions of this type
which are  Bose symmetric, invariant under the Euclidean group,
analytic in the flow parameter and  the momenta,
and satisfy the Osterwalder-Schrader growth condition.
\footnote{Christoph.Kopper@polytechnique.edu,}
\footnote{Postal Adress : Route de Saclay, 91128 Palaiseau, France}.
\footnote{Orcid-ID 0000-0003-2862-1864}

\end{abstract}

\vspace{1.5cm}


\newpage

\section{ Introduction }
\label{sec:intro}

Scalar $\,\vp^4_4$-theory plays a prominent role
in quantum field theory. In the standard model of particle 
physics it describes (part of) the Higgs sector of this model.
In the analysis of critical phenomena symmetric (one component) 
critical $\,\varphi^4_4$-theory belongs to
the universality class of the $d=4$ Ising model. Apart from direct physical 
applications, massive  $\,\varphi^4_4$-theory 
also serves as the simplest theoretical
framework to expose perturbation theory of
quantum field theories, preserving   its  generic complexity
including  the renormalisation problem, 
but avoiding  additional difficulties,
as e.g. those stemming from the implementation of gauge symmetry 
in the quantised theory or from infrared problems. 

Unfortunately, however, there is strong evidence 
that the model cannot be used to
define an interacting quantum field theory beyond perturbation
theory. In fact rigorous work shows that the continuum limits 
of lattice-regularized versions of  this model are 
 trivial, 
i.e. noninteracting field theories
~\cite{Froehlich82,Aizenman82,FFS,ADC2021}
under quite general assumptions. The triviality theorems
are conceptually  related to the fact that in 
the perturbative analysis, the effective
(so-called running) coupling constant of the model increases with
energy and even has a singularity at some finite energy, 
so that at high energy perturbation theory is no more 
reliable for calculations in physics.   

In this respect arises the question whether the modification
of the bare interaction of $\,\vp^4_4$-theory
could change the situation. In a previous paper ~\cite{Kopper} 
we analysed this question in the  simplified framework of the mean field
limit. 
We showed that bare actions enlarged by an 
infinite number of 
terms, individually  irrelevant in the sense of the renormalisation group,
  can give rise to asymptotically free solutions of the 
flow equations (FEs). 
In  ~\cite{Kopper} we studied the renormalisation group (RG)
flow in terms of a
logarithmic energy scale $\mu\,$, the value  
$\,\mu=0\,$ corresponding to the bare action and the value 
$\,\mu_{\textrm max} \gg 1\,$ 
to the fully integrated physical action. 
The limit $\mu_{\textrm max} \to \infty\,$ 
 corresponds to the removal of  the UV cutoff. 
We  proved the following  \textit{Theorem} :\\
\textit{There exist nonpolynomial mean field bare  actions $L_0(\bar \vp)\,$ 
  which, when restricted to finite volume,
  are uniformly bounded from below
and locally analytic in the 
(constant) mean field variable $\bar \vp \in \mathbb{R}\,$,
 such that the moments 
$A_n(\mu)$ of the effective actions $L_{\mu}(\bar \vp)$,
 as functions of the flow parameter $\,\mu\,$,
 $0 \le \mu \le \mu_{\textrm max}\,$, 
exist as locally analytic  solutions of the system of FEs,  
and have  nontrivial  ultraviolet limits for
$\,\mu_{\textrm max} \to \infty\,$.
In particular there exist  
asymptotically free bare actions of this type for which the bare couplings 
$\,A_n(0)\,, \ n \ge 4\,$, vanish logarithmically in the ultraviolet limit.}
\\[.1cm]      
This result was   unexpected, but not contradictory to present knowledge 
on mean field  $\,\vp^4_4$-theory. In particular 
we discussed its compatibility with the seminal
work of Felder ~\cite{Felder} on the subject.   
The mean field approximation  is closely
related to the so-called hierarchical models for which a rigorous  
fixed point analysis has been performed in ~\cite{GK}. 
These fixed points are 
 infrared fixed points
relevant to the analysis of critical phenomena 
~\cite{WF72}. Our results concern the Gaussian  ultraviolet        
fixed point characteristic of UV asymptotic freedom. 
We also checked consistency of our method with the above mentioned
triviality results by imposing bare actions of the strict 
$\vp^4\,$ type. In this case the triviality of mean field   
$\vp^4\,$-theory is recovered, as was first shown for sufficiently small
bare couplings in ~\cite{Kopper}
and then proven for arbitrary bare couplings 
~\cite{KW}, where the result was also extended to
$O(N)$-models and uniqueness of the trivial solution
for given bare couplings was established.

The differential FEs of 
the renormalisation group for the Schwinger functions
constitute the basic tool for this construction.
They are a consequence of the functional differential FE 
for the effective action  of the field theory considered.
The founding ideas go back to Wilson
\cite{Wilson71}. The differential form  appears
first in a paper by Wegner and Houghton ~\cite{Wegner72}. When
expanded in moments it becomes an infinite system 
of differential equations for the connected
free propagator amputated Schwinger functions (CAS) of the theory. 
 Polchinski  ~\cite{Polchinski84} observed that
when expanding these functions order by order in the number of loops, 
there is an airtight inductive scheme  permitting 
to control the perturbative functions and to establish 
renormalisability. In  subsequent
papers Polchinski's result was implemented in a 
rigorous path integral framework ~\cite{Hurd:1989},  
including physical renormalisation conditions
  ~\cite{Keller:1990ej, Keller:1991bz},
and later on extended to basically all 
quantum field theories of physical interest, in particular
to nonabelian gauge theories ~\cite{EGK, FroebH}.
For an early review see ~\cite{Muller:2002}.

Traditionally methods  stemming from statistical 
physics, like cluster and Mayer expansions 
~\cite{MayerMontroll, Brydges, Glimm, Riv:PCR, BBS}, 
are used for a rigorous nonperturbative analysis of  
(regularised) path integrals.
They stay technically relatively straightforward when applied to
 superrenormalisable field theory models like massive
$\vp^4_2\,$ and $\vp^4_3\,$ theory, or to models with short-range 
correlations in statistical physics. 
For a general introduction to these methods 
see ~\cite{Glimm, Simon}. 
The detailed review ~\cite{BBS}
shows the state of the art concerning 
theories which are not superrenormalisable or  have long range 
correlations. It mainly adresses the statistical physics 
context, focussed on infrared fixed points. 
Important progress has been made over the last decades,
but the review nevertheless  shows
that the subject is still highly technical
and intricate\footnote{Note that the main part of the review 
adresses the hierarchical approximation. The modifications required
for the extension to Euclidean field theory are commented in an
appendix referring to original papers.}.
The differential FEs  have not often been 
employed in a rigorous nonperturbative  context. 
Notable exceptions are the following. In  ~\cite{BK} the FEs
are integrated over a 
finite interval for bounded and analytic initial data.
In a purely fermionic context the FEs have been used recently
to construct nonperturbatively a weakly interacting fixed point
~\cite{Gr} of a Gross-Neveu type model. And the FEs
in integrated form  were also  employed  to construct the effective potential
and the Schwinger functions of the two-dimensional Gross-Neveu model
in the preprint ~\cite{Du}. In a recent paper ~\cite{OSa} an elegant
tree expansion applied to Polchinski's equation
transformed into an integral equation,  has been used to prove
convergence results for an expansion w.r.t. the (bare) interaction
in a fermionic context.

The fact that the mathematical structure of 
the system of FEs for the $\,n$-point functions
of Polchinski, as we use it,
is not straightforwardly adapted to such an analysis,
is related to its dependence on the index $\,n\,$. 
Since Polchinski's inductive 
scheme is not viable beyond perturbation theory,
one might be tempted to look for solutions obtained 
by  iteratively integrating the FEs. But it turns out that 
the $\,n\,$-dependence of the system generally makes diverge such iterative
constructions.
Therefore  the strategy followed in ~\cite{Kopper}
was to  consider mean field two-point  functions,
locally analytic (or at least smooth) w.r.t. the flow parameter, 
and to inductively construct higher  $\,n$-point functions from the 
lower ones. Since the $\,n+2$-point function appears in the flow equations
as a function of the $n'$-point functions and their derivatives
w.r.t. the flow parameter, where $\,n' \le n\,$,  it is possible
to set up an inductive scheme in $n$ and the number of derivatives
in order to construct all $n$-point functions. 
For solutions locally analytic 
w.r.t. the flow parameter, uniqueness of the solutions 
can be established.  
This procedure differs from the standard way of
analysing dynamical systems, which are integrated starting from
explicit boundary conditions. Here
the (in a first moment implicit) boundary conditions are obtained
as a consequence
of the choice of the two-point function. 
For locally analytic  two-point functions the constructed
solutions of the FEs for the $n$-point functions
inherit local analyticity, and they are
unique in this regularity class.
In this way we constructed mean field solutions  
~\cite{Kopper} 
for  bare actions (i.e. boundary 
conditions) which are nonpolynomial in the mean field variable
and analytic in the flow parameter.  

Quite generally, the bare actions considered in 
quantum field theory are  local polynomials
in the field variable(s) and its (their) derivatives.
The perturbative calculations performed on that basis
in theories like QED,  are highly successful and put stringent bounds 
 on higher order irrelevant terms in the bare action.
On the other hand bare actions containing only a few local monomials,
appear unnatural 
from the point of view of the Wilson renormalisation group (RG)
 since they are related to an impressive fine tuning procedure.  
Namely starting from such a bare action
and integrating out degrees of freedom (in whatever small 
a range),  one immediately obtains 
an effective action containing 
nonlocal monomials of any degree in the fields
(as far as they are allowed by the symmetries of the theory).
Since the UV cutoff has no intrinsic meaning, this new
action could as well have been taken as a starting point.
The fine tuning consists in arranging 
the renormalisation group trajectory  
such that at a specific value of the cutoff (the one chosen initially)
the infinite number of all
higher order moments of $\,L_0$ are forced to pass through $0$ each,
and this for all momentum or position arguments.
The way in which the Wilson RG
is normally exposed heuristically,
says that all these fine-tuned terms do not really matter
for the fixed point analysis of the RG, since any local 
monomial in the field and its derivatives of scaling dimension higher 
than four is RG irrelevant. But this argument does not address the 
status of  infinite sums of such  monomials, or even nonlocal
versions thereof. We note that the question of nonpolynomial 
actions has been adressed in a theoretical physics context 
in particular by Halpern and Huang   ~\cite{HH}, and refuted on
the basis of a more thorough analysis by Morris  and collaborators
~\cite{Morris}.
In both cases the analysis was based on a linearisation of the
mean field flow equations close to a Gaussian fixed point. 
But the question of the existence of global solutions of the
flow equations is beyond the scope of the linearised equations
anyway, as is clear from ~\cite{Felder} and also from
our subsequent analysis. 
So we cannot take any input from these papers, 
even if there is an interesting analysis of Kummer's equation in
 ~\cite{Morris}. 

In the present paper we extend the analysis from ~\cite{Kopper}
beyond  the mean field limit.  
Our point is to show  again that nonpolynomial bare  actions
can lead to essentially different, in particular UV
asymptotically free,  solutions of the FEs 
 which escape the  triviality
statement. 
The inductive scheme used to construct these solutions is the same 
as in the mean field approximation used in ~\cite{Kopper}. 
The main new difficulty
is due to the fact that the $n+2$-point function appears in the
FEs only in integrated form. Requiring  local analyticity  
w.r.t. the flow parameter, and  now also with respect to the 
four-momenta,  makes it possible to relate 
the ambiguity in choosing the integrand to the boundary conditions
imposed for the bare action, different choices leading to 
different bare actions. The choice
which we adopt in our construction is then dictated
by the principle that locality should be respected
``as much as possible''.

Here we do not address certain important farther reaching 
questions concerning the physical relevance of these solutions
and their status w.r.t. the path integral.
Aspects which remain to be verified (or falsified)
are the existence and control
of the bare action, in particular whether it can be chosen 
bounded from below, the full set of the 
Osterwalder-Schrader axioms ~\cite{OS} and  possibly
also the direct explicit
construction in Minkowski space,
or the unitarity of the corresponding $S$-Matrix if any.
Our construction satisfies (as will be shown) three
of the five Osterwalder-Schrader axioms, namely Euclidean
invariance, Bose symmetry and the Osterwalder-Schrader growth
condition. Clustering should follow straightforwardly,
though we do not adress it here. Osterwalder-Schrader
positivity is not evident. Since our construction starts
from the two-point function, a first step might be  to identify
two-functions which satisfy the K\"all\'en-Lehmann representation.
Another open question is the verification of the
triviality of scalar field theory for the minimal  $\vp_4^4$
bare action within this framework.

This paper is organized as follows\,:
In section \ref{sec:FEs} we recall the FE framework.
In section \ref{linear} we introduce the vector spaces 
suited for the inductive construction of the solutions
of the FEs. In particular we take into account 
Bose symmetry and translation invariance.
In section \ref{sec:analpre} we present our technique
of how to construct in this context
$(n+2)$-point functions from  $n$-point functions, 
which satisfy convenient $L^\infty$-bounds. 
In section \ref{bounds} we inductively 
bound solutions of the FEs starting from a suitable class
of two-point functions. The bounds obtained imply the
Osterwalder-Schrader growth condition.
In section \ref{smgs} we prove the existence of
nontrivial  UV asymptotically free  global solutions 
for the system of $n$-point functions, which are locally analytic w.r.t.
the flow parameter and the four-momenta,
and under these conditions unique for a given bare action
in the considered class.


\section{ The flow equations }
\label{sec:FEs}
We consider a (one-component) 
self-interacting scalar field  on four dimensional 
Euclidean space. We adopt the renormalisation group (RG) 
flow equation (FE) framework. 
In the following  we will give a brief review of the general formalism and 
define the  objects of interest for the purpose of this paper. 
See ~\cite{Muller:2002}  for a more 
comprehensive review of the flow equation approach within our context.
Formulating our theory with ultraviolet (UV)
 and infrared (IR) cutoff in the standard path integral 
formalism requires two main ingredients:
\begin{enumerate}
\item We define the regularised propagator or covariance
  in  momentum space as
\ben
\!\!\!\!\!\!\!\!\!\!\!C^{\al_{0},\al}(p;m)=\frac{e^{-\al_0(p^{2}+m^2)}
-e^{-\al (p^{2}+m^2)}}{p^{2}+m^2}\ , \ \ 
\, p = (p^1,\ldots,p^4) \in \mr^4\ , \quad 0 \le \alo \le \al < \infty\ .
\label{cov}
\een
Upon removal of the cutoffs, i.e. in the limits 
$\al_0\to 0$ (UV), $\al\to \infty$ (IR), we indeed recover the free
propagator 
$\frac{1}{p^{2}+m^2}\,$.
For the Fourier transform, we use the convention
\ben
\ti f(x) = \int_p   f(p)\, \e^{ipx}\ \  \mbox{ using the shorthand }\ \
\int_p  := \int \frac{\d^4p}{(2\pi)^4}
\een
so that in position space
\eq
\ti {C}^{\al_0,\al}(x-y;m) = 
\int_p  e^{ip(x-y)}\ C^{\al_0,\al}(p;m) \ . 
\eqe
\item The bare interaction Lagrangian $\,L_{0}(\vp)\,$ is supposed
to be well-defined for fields $\,\vp\,\in {\textrm supp}\ \mu^{\al,\al_0}\,$.
Here   $\mu^{\al,\al_0}$
is the normalised Gaussian measure with covariance
\eqref{cov}
\footnote{
See the Appendix to Part I of~\cite{Glimm} for mathematical details 
about Gaussian functional integrals.}. 
The \emph{basic field} $\varphi $ is  assumed to be in the 
support  of $\mu^{\al,\al_0}$ and thus to be smooth in position space
\footnote{We do not adress questions related to the infinite volume
limit in this paper since we will state on the connected amputated 
Schwinger functions for which this limit is not problematic.
For questions related to the effective action itself the limit has to
be analysed. See  ~\cite{Muller:2002} for more details.}.
\end{enumerate}
The effective action at scale $\,\alpha\,$
is then given by
\ben\label{CAGdef}
-L^{\al_0,\al}(\vp)  := \, \log \biggl[
\int \d\mu^{\al_0,\al}(\phi)  \exp \Big(- L_0 (\vp+\phi)\Big)\biggr]
-  \log  Z^{\al_0,\al} \ .
\een
The normalisation factor $Z^{\al,\al_0}$
is chosen so that $\,L^{\al_0,\al} (0)=0\,$.
The moments $\,\bar\L^{\al_0,\al}_{n}\,$ of  $\,L^{\al_0,\al} (\vp)\,$
at $\,\vp\equiv 0\,$, obtained from the expansion
\ben\label{genfunc}
L^{\al_0,\al}(\varphi) := \sum_{n\in 2 \mathbb{N}}
\int \frac{\d^4 p_1}{(2\pi)^{4}} \dots 
\frac{\d^4 p_n}{(2\pi)^{4}}\ \bar{\L}^{\al_0,\al}_{n}(p_1, \ldots, p_n)
\,
\hat\varphi(p_1) \cdots \hat\varphi(p_n)\ ,
\een
are the connected amputated Schwinger functions (CAS)
with propagators \eqref{cov}. 
No statement is made here about the
convergence of the series \eqref{genfunc}. 
Translation invariance of the CAS  in 
position space implies that
$\bar{\L}^{\al_0,\al}_{n}(p_1, \dots, p_n)$ is supported
at $p_1+\ldots+p_n=0$ 
(momentum conservation), and thus only depends on $n-1$ 
independent four momenta. In fact Fourier transformation
shows that we can write
\ben
\label{deltaCAG}
\bar{\L}^{\al_0,\al}_{n}(p_1, \dots, p_n)\,= \,
\delta^{4}{(\sum_{i=1}^n p_i)}\ \L^{\al_0,\al}_{n}(p_1, \dots, p_n) \ .
\een
The functions 
$\bar{\L}^{\al_0,\al}_{n}(p_1, \dots, p_n)\,$ are fully 
symmetric under permutation of {\textit all} $\,p_{1},\ldots, p_{n}\,$.
Note  that the previous definitions imply that the free
(amputated) two-point function is not included in the function 
$\L_2(p_1,p_2)\,$, but  has been absorbed in the Gaussian measure.

When deriving \eqref{CAGdef} w.r.t. $\al\,$, we obtain
the functional FE of the RG\,:
\ben
\pa_{\al}L^{\al_0,\al}(\varphi)\,=\,
\frac12 \langle \de_{\vp} , \dot {\ti C}^\al \de_{\vp} \rangle L^{\al_0,\al}(\vp)
\,-\,
\frac12 \langle \de_{\vp} L^{\al_0,\al}(\vp), 
\dot {\ti C}^\al \de_{\vp} L^{\al_0,\al}(\vp)\rangle\  .
\label{fe}
\een
Here $ \langle\ , \  \rangle $ denotes the scalar product in
$L^2(\mr^4, \d^4 x)\,$, and $\,\dot {\ti C}^\al$ is short for 
\ben
\dot {\ti C}^\al(x-y;m)\,=\, \int_p  e^{ip(x-y)}\ \dot C^{\al}(p;m)
\ ,\quad \dot C^\al(p;m) 
\,=\, \frac{\pa}{\pa \al} C^{\al_0,\al}(p;m)\,=\,
e^{-\al(p^2+m^2)}\ . 
\label{cdot}
\een
When expanding \eqref{fe} in moments, we get the expanded FEs
 for the CAS in momentum space 
\ben
\label{CAGFEsexpand}
\begin{split}
&\partial_{\al}{\L}^{\al_0,\al}_{n}(p_{1},\ldots,p_{n})
  = \genfrac(){0pt}{1}{n+2}{2}\,
\int_{p} \dot{C}^{\al}(p;m)\,\L^{\al_0,\al}_{n+2}(p_{1},\ldots,p_{n}, p, -p)\\
&-\frac{1}{2}\sum_{n_{1}+n_{2}=n+2} n_{1}n_{2}\   
\mathbb{S}\, \left[  \L^{\al_0,\al}_{n_{1}}(  p_{1},\ldots,p_{n_{1}-1}, p')\,
  \dot{C}^{\al}(p';m)\,   \L^{\al_0,\al}_{n_{2}}(-p',  p_{n_{1}},\ldots,p_{n}) 
\right]\, ,
\end{split}
\een
with $\,p'=p_{n_{1}}+\ldots+p_{n}=-p_{1} -\ldots -p_{n_{1}-1}\,$.
In \eqref{CAGFEsexpand}
$\,\mathbb{S}\,$ {\textit is the symmetrisation operator} acting linearly
on sums of 
(products of) functions, depending on
the four-momenta $(p_{1},\ldots, p_{n})\,$,
by taking the symmetrised mean value
\ben
\mathbb{S}[f(p_{1},\ldots, p_{n})]\,:=
 \frac{1}{n !}  \sum_{\si \in S_n}
f(p_{\si(1)},\ldots,p_{\si(n)})\ . 
\label{mS}
\een
Since the functions $\,{\L}^{\al_0,\al}_{n}(p_{1},\ldots,p_{n})\,$
are already symmetrised, we find in this case  
\[
\mathbb{S} [\L^{\al_0,\al}(p_{1},\ldots,p_{n_{1}})\,
\L^{\al_0,\al}(p_{n_1+1},\ldots,p_{n})]\,=
\]
\ben
\genfrac(){0pt}{1}{n}{n_1}^{-1}\,
\sum_{\si \in S_n}{\!\!\! '} 
\L^{\al_0,\al}(p_{\si(1)},\ldots,p_{\si(n_{1})})\,
\L^{\al_0,\al}(p_{\si(n_1+1)},\ldots,p_{\si(n)})\ , 
\label{mSs}
\een
where $\sum\,'$ goes over all $\si \in S_n\,$ such that
$\si(1)<\si(2) <\ldots<\si(n_{1})$  and $\si(n_{1})<\ldots<\si(n)\,$,
$\,S_n\,$ being the symmetric group  of $\,n\,$ elements.
These restricted permutations are also called shuffles.
For even scalar theories
only even moments (i.e. for even values of $\,n,\,n_1,\,n_2\,$) 
will be nonvanishing, 
due to the symmetry $\varphi\to-\varphi\,$. The infinite system of 
equations \eqref{CAGFEsexpand}  constitutes
an  infinite dimensional nonlinear dynamical system.
\vspace{.2cm}
{\remark 
Since we are interested in the ultraviolet behaviour 
of the theory, we will simplify the presentation as we did in ~\cite{Kopper},
by setting  $\,m=0\,$ and limiting
$\al\,$ to the finite interval $[\al_0, 1]\,$.
Then $\,\al_{\textrm max} =1\,$ takes the role of the  infrared
regulator $\frac{1}{m^2}\,$. 
This technical simplification can
be avoided by considering a differently  parametrised flow of the real
massive propapagtor as in ~\cite{KW}. There it was shown that in the
mean field case the conclusions stay unaltered.
We do not see obstructions of principle for extending this statement
to the full system of FEs. But explicit verification would of course
be highly welcome.}\\[.2cm]
We use the shorthands
\ben
C^{\al_{0},\al}(p):=\,
C^{\al_{0},\al}(p;0)\ , \quad 
\dot C^{\al}(p):=\,
\dot C^{\al}(p;0) \,=\, 
e^{-\al \,p^2}\ ,
\label{cov0}
\een
\ben 
\dot \cC^{\al}(p) :=\, c\ \al^2\,\dot{C}^{\al}(p) 
\quad
\mbox{with} \quad c=16\, \pi^2\ \quad
\mbox{so that}\quad \
\int_p \dot \C^\al(p)\,=\,1\ .
\label{norm}
\een

\vspace{.1cm}

\noindent
To factor out the basic scaling behaviour w.r.t. $\,\al\,$ and $\,c\,$
we define 
\eq
   {\ell}^{(\al)}_{n}(\mu;p_{1},\ldots,p_{n})
   \,:=\,
   n\  c^{1-\frac{n}{2}}\ \al^{2-n/2}\
   {\L}^{\al_0,\al}_{n}(p_{1},\ldots,p_{n})
\label{scaling1}
\eqe

with 
\ben
\mu\,=\, \ln \frac{\al}{\al_0} \quad \Leftrightarrow \quad
 \frac{\al}{\al_0} \,=\, e^{\mu}\ .
\label{mu}
 \een
We suppressed the index $\,\al_0$ in $\,{\ell}^{(\al)}_{n}\,$, assuming 
it to be fixed. 
The system  \eqref{CAGFEsexpand} can then be rewritten 
\ben
\label{fescale}
\begin{split}
&\int_{p} \dot{\C}^\al(p)\,\ell^{(\al)}_{n+2}(\mu;p_{1},\ldots,p_{n}, p, -p)
\\
&=\ \frac{1}{n+1} \sum_{n_{1}+n_{2}=n+2}\!\! 
\mathbb{S}\, \left[  \ell^{(\al)}_{n_{1}}( \mu; p_{1},\ldots, p')\,
\dot{C}^{\al}(p')\,   \ell^{(\al)}_{n_{2}}(\mu;-p', \ldots,p_{n}) 
\right]\\\
&
\,  + \ \frac{n-4}{n(n+1)}\,{\ell}^{(\al)}_{n}(\mu;p_{1},\ldots,p_{n})+\,
\frac{2}{n(n+1)}\,\partial_{\mu}\,{\ell}^{(\al)}_{n}(\mu;p_{1},\ldots,p_{n})
 \  .
\end{split}
\een
\vspace{.2cm}
\noindent
An inductive analysis of the FEs requires also to study their
derivatives w.r.t. the flow parameter $\,\mu\,$. These derivatives
are to be taken at fixed $\,\vec p = (p_1,\ldots,p_n)\,$.
Acting with $\,l\,$ $\,\mu$-derivatives on \eqref{fescale}
we get
\ben
\label{fescalel}
\begin{split}
&\pa^{l}_{\mu}\, \int_{p} \dot{\C}^\al(p)
\,\ell_{n+2}(\mu;p_{1},\ldots,p_{n}, p, -p)\\
&=\,\frac{1}{n+1}\ 
\sum_{n_{1}+n_{2}=n+2\atop l_1+l_2+l_3=l} \Bigl({l\,\atop l_1\, l_2 \,l_3}\Bigr)\
\mathbb{S}\, \left[ [\pa^{l_1}_{\mu}\, 
\ell_{n_{1}}](\mu;p_{1},\ldots, p')\,
[\pa^{l_2}_{\mu}\, \dot{C}^{\al}](p')\, [\pa^{l_3}_{\mu}\,  
\ell_{n_{2}}](\mu;-p',\ldots,p_{n}) 
\right]\\
&+\,
\frac{n-4}{n(n+1)}\,\pa^{l}_{\mu}\,{\ell}_{n}(\mu;p_{1},\ldots,p_{n})\,+\,
\frac{2}{n(n+1)}\,\pa^{l+1}_{\mu}\,{\ell}_{n}(\mu;p_{1},\ldots,p_{n})\ .
\end{split}
\een
The derivatives on the l.h.s can be expanded
\[
\pa^{l}_{\mu}\, \int_{p} \dot{\C}^\al(p)
\,\ell_{n+2}(\mu;\ldots,p, -p)\,=\,
\sum_{0 \le \la \le l} \Bigl( {l \atop \la} \Bigr)
\,\int_{p} (\pa^{l-\la}_{\mu}\dot{\C}^\al(p))  
\,(\pa^{\la}_{\mu}\,\ell_{n+2})(\mu; \ldots, p, -p)\ .
\]

\vspace{.3cm}
\noindent
It turns out to be more practical to work with the 
functions
\ben
\label{scaling2}
\ell_{n}(\mu;q_{1},\ldots,q_{n}):= \ell^{(\al)}_{n}(\mu;p_{1},\ldots,p_{n})\ ,
\ \ \mbox{ where }\ \ q_i:=\sqrt{\al}\,p_i\ .
\een
In terms of the $\ell_n(\mu;q_1,\ldots,q_n)\,$ the FEs \eqref{fescale}
can be written 
\ben
\label{cagfe}
\begin{split}
&\int_{q} \dot{\C}^1(q)
\,\ell_{n+2}(\mu;q_{1},\ldots,q_{n}, q, -q)\\
&\quad =\,\frac{1}{n+1}\ 
\sum_{n_{1}+n_{2}=n+2} 
 \mathbb{S}\, \left[
\ell_{n_{1}}(\mu;q_{1},\ldots, q')\,
\dot{C}^{1}(q')\,  
\ell_{n_{2}}(\mu;-q',\ldots,q_{n}) 
\right]\\
&\quad+\,
\frac{n-4}{n(n+1)}\,  {\ell}_{n}(\mu;q_1,\ldots,q_{n})\,+\,
\frac{2}{n(n+1)}\, \pa_\mu\Bigl|_{\{\vec p\}} {\ell}_{n}(\mu;q_{1},\ldots,q_{n})\ .
\end{split}
\een
When changing from the functions
\eqref{scaling1} to \eqref{scaling2}
we have to
note that 
\ben
\label{dmu}
\pa_\mu\Bigl|_{\vec p_n}
\ell_{n}(\mu;q_{1},\ldots,q_{n})\,=\,
\pa_\mu\Bigl|_{\vec q_n}{\ell}_{n}(\mu;q_{1},\ldots,q_{n}) \,+\,
\frac12  \sum_{i=1}^n \sum_{\nu=1}^4
q_{i,\nu}\,\pa_{q_{i,\nu}}\,{\ell}_{n}(\mu;q_{1},\ldots,q_{n}) \ .
\een
Subsequently we will work with the functions
\eqref{scaling2}, and then we will always understand that
\[
\pa_\mu\,:=\,\pa_\mu\Bigl|_{\vec q}\ .
\]
In view of \eqref{dmu} we introduce the shorthand
\ben
\label{shorthand1}
\frac{d}{d\mu} {\ell}_{n}(\mu;q_{1},\ldots,q_{n})\,:=\,
\partial_{\mu}\,{\ell}_{n}(\mu;q_{1},\ldots,q_{n})
\,+\,\frac12 \sum_{i=1}^n  \sum_{\nu=1}^4
q_{i,\nu}\,\pa_{q_{i,\nu}}\,{\ell}_{n}(\mu;q_{1},\ldots,q_{n})\ .
\een
In order to describe how to act with
successive derivatives on \eqref{cagfe}
we introduce a multi-index to be  written generally
$\,w\,$\footnote{More precisely we should write
$w_{n+1}\,$. But this index will be generally
clear from the context.}.
We  set
\ben
w \in \mathbb{N}_0^{n+1}\ ,\quad  w =(w^{(0)},w^{(1)},\ldots,w^{(n)}) \ , \ \
w^{(i)}  \in \mathbb{N}_0\ , \ \ \
 |w|\,:=\, \sum w^{(i)} \ , \ \
w !:= \prod w^{(i)}  ! \ ,
\label{w1}
\een
and define
\ben
\pa^w \ell_n(\mu; q_1,\ldots,q_n) \,:=\,
\pa^{w^{(0)} }_\mu\ \prod_{i=1}^n \left(\frac12 \sum_{\nu=1}^4\,q_{i,\nu}\,
\pa_{q_{i,\nu}}\right)^{\! w^{(i)} }
  \ell_n(\mu; q_1,\ldots,q_n)\ .
\label{w2}
\een
Then we can also write the FEs \eqref{cagfe} derived by $\pa^w$
\ben
\label{cagfew}
\begin{split}
&\int_{q} \dot{\C}^1(q) \pa^w 
\ell_{n+2}(\mu;q_{1},\ldots,q_{n}, q, -q)\\
&=\,\frac{1}{n+1}\ 
\sum_{n_{1}+n_{2}=n+2\atop w_1+w_2+w_3=w} c_{w_1,w_2,w_3}
\, \mathbb{S}\, \left[
\pa^{w_1} \ell_{n_{1}}(\mu;q_{1},\ldots, q')\,
\pa^{w_3} \dot{C}^{1}(q')\,  
\pa^{w_2} \ell_{n_{2}}(\mu;-q',\ldots,q_{n}) 
\right]\\
&+\,
  \frac{n-4}{n(n+1)}\, \pa^w
  \, {\ell}_{n}(\mu;q_1,\ldots,q_{n})\,+\,
\frac{2}{n(n+1)}\,\frac{d}{d \mu}\pa^{w}
\,{\ell}_{n}(\mu;q_{1},\ldots,q_{n})\ ,
\end{split}
\een
where the combinatoric factor $\,c_{w_1,w_2,w_3}\,$ is given by
\ben
c_{w_1,w_2,w_3}:= \frac{w!}{w_1! w_2! w_3!}\ , \quad w=w_1+w_2+w_3\ .
\label{cw}
\een


\section{Linear spaces  of functions }
\label{linear}

We define for $\al > 0$
\ben
{\cal H}^\al_n\,=\, L^2(\mathbb{R}^{4n},\,\d\mu^\al(p_1,\ldots,p_n))\ ,
\ \  p_i \in\mathbb{R}^{4}\ ,\ \ n \in \mn\ ;\quad
{\cal H}^\al_0\,=\, \mathbb{R}\ .
\een
So the spaces $\,{\cal H}^\al_n\,$ are    the  Hilbert spaces of real
functions on $\,\mathbb{R}^{4n}\,$, square integrable
with respect to the normalised measure   (see \eqref{norm})
\ben
d\mu^\al(p_1,\ldots,p_n)\,=\,  \dot {\cal C}^\al(p_1)\ldots 
\dot {\cal C}^\al(p_n)\ \frac{\d^4p_1}{(2\pi)^4}\ldots
\frac{\d^4p_n}{(2\pi)^4}\ .
\label{mas}
\een
The scalar product is denoted by $\, (\ ,\ )^\al\,$. 
We consider  orthonormal (ON) bases of $\,{\cal H}^\al_1\,$
with basis functions
$\,\psi^\al_k(p)\, ,\ k \in \mn_0^4\,$, with the conditions that
\ben
\psi^\al_0(p)\,=\, 1\!\!\!1(p)\,\equiv \, 1 \ , \ \
\mbox{where}\ \   0 \ \  \mbox{stands for}\ \ (0,0,0,0)\ ,\quad
\psi^\al_k(-p)\,=\,\pm \psi^\al_k(p)\ .
\label{1par}
\een 
 We also write for $ n \ge 1$
\[
\1^{n}(p_1,\ldots,p_n)\,=\,\1(p_1)\ldots\1(p_n)\ .
\]
Allowed sets of basis  functions would be e.g. normalised Hermite polynomials,
or normalised products of Laguerre polynomials and hyperspherical harmonics.
The functions
\ben
\psi^\al_{(k_1,\ldots,k_n)}(p_1,\ldots,p_n)\,:=\,
\psi^\al_{k_1}(p_1)\,\ldots\, \psi^\al_{k_n}(p_n)\ , \ \ k_i \,\in\,
\mn_0^4\ \, ,
\label{onb}
\een
constitute an ON basis of  $\,{\cal H}^\al_n\,$. We have
\[
\psi^\al_{(0, \ldots,0)}(p_1,\ldots,p_n)\,=\,
\1^{n}(p_1,\ldots,p_n)\,\equiv \, 1\ .
\] 
From orthogonality $\,(\psi_k, \1)^\al \,=\, 0\,$ for $\, k \neq (0,0,0,0)\,$,
we obviously have
\[
\int \psi^\al_k(p)\ d\mu^\al(p) =\, 0 
\quad \forall k \,\neq\, (0,0,0,0)\ ,
\]
and more generally
\ben
\!\!
\int \psi^\al_{(k_1,\ldots,k_n) }(p_1,\ldots,p_n)\,
d\mu^\al(p_1,\ldots,p_n) = 0 \ \ 
 \mbox{if at least one }\,   k_i \, \mbox{ does not equal } (0,0,0,0)\ .  
\label{ortho}
\een 
Our notations are adapted to the fact that we are interested
in functions  invariant under the Euclidean group.
We will often
consider functions $f\in {\cal H}^\al_n\,$ in momentum space satisfying
\[
f(p_1,\dots,p_n) \,=\, f(O p_1,\dots, O p_n)\quad 
\forall \ O \in O(4)\ ,
\]    
where $\,O(4)\,$ is the orthogonal group acting on $\,\mr^4\,$.


\subsection{Bose symmetry}
\label{Bose}

 {\definition
The symmetric  Hilbert subspaces $\,{\cal S}^\al_n\,
\subset \,{\cal H}^\al_n\,$ 
 are defined through
\ben
{\cal S}^\al_n\,:=\, \{ \psi^\al_n\in {\cal H}^\al_n\,\Big|\,  
\psi^\al_n(p_1,\ldots, p_n)=
\psi^\al_n(p_{\si(1)}, \,\ldots\,  ,p_{\si(n)})\   \forall \si \in S_n \}\ .
\label{defS}
\een
We also set
\ben
{\cal S}^{\al,\infty}_n\,:=\, \{ \psi \in {\cal S}^\al_{n}\ |\   
\sup_{\vec p} |\psi(\vec p)| < \infty\} \ ,\quad 
{\cal S}^\al_0\,= \, \mathbb{R}\ ,\quad
{\cal S}^\al_n\,=\ \,{\cal H}^\al_n\,=\,\emptyset\ 
\ \forall\ n \in -\mathbb{N}\ .
\een
\label{ddefS}
}
The subspaces 
$\,{\cal S}^{\al,\infty}_{n}\,$ are dense in $\,{\cal S}^\al_{n}\,$
w.r.t. to the $L^2$-norm. 
\noindent
Mutually orthogonal basis functions of $\,{\cal S}_n\,$ 
can be obtained from the
ON basis functions $\,\psi^\al_{k_{1}}(p_1)\,\ldots\,\psi^\al_{k_{n}}(p_n)\,$ 
of $\, {\cal H}^\al_n\,$ by  restricting to symmetrised  functions 
\ben
\psi^\al_{(k_1,\ldots, k_n)}(p_1,\ldots, p_n):=\,
\sum_{\si} \psi^\al_{k_{\si(1)}}(p_1)\,\ldots\, 
\psi^\al_{k_{\si(n)}}(p_n)\ .
\label{sbasis}
\een
To normalise them to $1\,$, one has to divide by a normalization factor 
\ben
{\cal N}(\vec k)\,=\, \sqrt{n!}\ \prod_{i=1}^{j} \sqrt{v _i!} \ \, . 
\label{nfr}
\een
Here we suppose that
the four-indices $\,k_1,\,\ldots, k_n\,$
can be grouped into $\,j\,$
blocks  of multiplicities $\,v _i \,$ of equal indices,
which implies that $\,\sum v _i \,=\,n\,$.
For a given choice of normalised $\psi^\al_{k_i}$
we denote the corresponding 
normalised basis function as 
\ben
\phi^\al_{(k_1,\ldots, k_n)}(p_1,\ldots, p_n):=\,
\frac{1}{ {\cal N} (\vec k)}\,\psi^\al_{(k_1,\ldots, k_n)}(p_1,\ldots, p_n)\ .
\label{nsbasis}
\een

\vspace{.2cm}
 {\definition
For $\, 0 \le \nu \le n\,$ the linear subspaces  
$\,{\cal H}^\al_{v,n-v}\,$ of $\,{\cal H}^\al_n\,$, and  
$\,{\cal S}^\al_{v,n-v}\,$ of  
$\,{\cal S}^\al_{n}\,$, are defined according to the multiplicity 
of the basis functions $\,\1(p)\,$. We set 
\ben
  {\cal H}^{\al,0}_{v,n-v} :=
\label{Hvnv}
  \een
  \[
{\textrm span} \{\,\psi^\al_{k_{1}}(p_{\si(1)})\ldots 
\psi^\al_{k_{v}}(p_{\si(v)})\, \1(p_{\si(v+1)})\ldots \1(p_{\si(n)})\, \Bigr| 
\  k_i \in \mn^4_0\,, \ 
k_i \neq (0,0,0,0)\,,\ \si \in S_n\, \}\ .
\]
The $L^2$-closures are written as 
\ben
{\cal H}^\al_{v,n-v} := \overline{{\cal H}^{\al,0}}_{v,n-v}\ . 
\label{hnuclosed}
\een
Analogously we set (remember \eqref{mS})
\ben
{\cal S}^{\al,0}_{v,n-v} := 
{\textrm span} \{\mathbb{S}[\psi^\al_{k_{1}}(p_1)\ldots \psi^\al_{k_{v}}(p_{v})\, 
\1(p_{v+1})\ldots \1(p_{n})]\, \Bigr| \  k_i \in \mn^4_0\,, \ 
k_i \neq (0,0,0,0)\,\} \ ,
\label{snubasis}
\een
and we again write for the $L^2$-closures
\ben
{\cal S}^\al_{v,n-v} :=\overline{ {\cal S}^{\al,0}_{v,n-v}}\ \, .
\een
\label{vn-v}
}
The subspaces $\,{\cal S}^\al_{v,n-v}\,$ are mutually orthogonal,
and $\,{\cal H}_n\,$, $\,{\cal S}_n\,$ have orthogonal decompositions~:
\ben
{\cal H}^\al_n\,=\,\bigoplus_{v =0}^{n}{\cal H}^\al_{v,n-v }\ , \quad
{\cal S}^\al_n\,=\,\bigoplus_{v =0}^{n}{\cal S}^\al_{v,n-v }\ ,\quad
\mbox{where}\quad  {\cal S}^\al_{v,n-v } \subset {\cal H}^\al_{v,n-v }\ .
\label{sn}
\een
Therefore a function $\psi_n \in {\cal S}^\al_{n}\,$
has a unique orthogonal 
decomposition
\ben
\psi_n\, =\, \sum_{v=0}^{n}  \psi^\al_{v,n-v}\ , \quad   
\psi^\al_{v,n-v} \, =\, 
\ms\, \Bigl[\,\psi^{\al}_v\; \1^{n-v}\,\Bigr]\,\in\, {\cal S}^\al_{v,n-v} \ \ 
\mbox{such that } \ \ \psi^{\al}_v \in {\cal S}^\al_{v,0}\ .
\label{decos}
\een
Then if
$||\psi^{\al}_v||_{\infty} \le C_v\,$ for suitable 
$\,C_v \ge 0\,$, we have
$\,||\psi_n||_{\infty} \,\le\, \sum_{v=0}^n C_v\,$.

\vspace{.2cm}

{\lemma 
\label{lms} 
The symmetrisation operator $\,\ms\,$ defined in 
\eqref{mS}, \eqref{mSs},
has the following properties~:\\[.1cm]
a) It acts linearly on the spaces 
$\,\cH^\al_n\,$. Its  $\,L^2$-norm equals $\,1\,$.\\[.1cm] 
b) It maps $\,\cH^\al_n\,$ onto  $\,\cS^\al_n\,$, 
and  $\,{\cal H}^\al_{v,n-v}\,$ onto  $\,\cS^\al_{v,n-v}\,$.\\[.1cm] 
c) $\,[\ms\,(f)](p_1,\ldots, p_n)\,=\,
[\ms\,(f)](p_{\si(1)},\ldots, p_{\si(n)}) \,$
for all permutations $\,\si\, \in S^\al_n\,$.\\[.1cm]  
d) For $\,f_i \in \,{\cal H}^\al_{n_i}\,$ we have
\ben
\ms\,\Bigl[\, (\ms f_1)\, f_2)\,\Bigr]\,=\,
\ms\,\Bigl[\,  f_1\, f_2 \,\Bigr]
\een
on $\,{\cal H}^\al_{n_1} \bigotimes {\cal H}^\al_{n_2}\,$.
}

\vspace{.2cm}

\noindent
{\textit{ Proof}}. All properties follow immediately
from the defnition.
The  $L^2$-norm of $\, \ms\,$ saturates 
for completely symmetric functions for which it acts as 
the identity. By the triangular inequality we generally have
\[
||\ms\,f||_2 \le ||f||_2\ .
\]
\hfill \proofSymbol

\vspace{.2cm}

\noindent

{\lemma
\label{lnfs} 
On the spaces ${\cal S}^\al_{v} \bigotimes {\cal S}^\al_{n-v}$ 
we define the operators  $\,\hms_{v,n-v}\,$
through
\ben
\hms_{v,n-v} \,[f_1\,f_2]\,=\,
\genfrac(){0pt}{1}{n}{v}\
\ms \,[f_1\,f_2]\quad \mbox{for }  \ f_1 \in{\cal S}^\al_{v}\, ,  \
f_2 \in {\cal S}^\al_{n-v}\ .
\label{hatms}
\een
An orthonormal basis of the space ${\cal S}^\al_{v,n-v}$ is given
by the functions
\ben
\vp^\al_{(k_1,\ldots,k_v)} :=
\genfrac(){0pt}{1}{n}{v}^{-1/2}\
\hms\,\bigl[\,\phi^\al_{(k_1,\ldots, k_v)}\,
\1^{n-v} \,\bigr] 
\; , \  k_i \in \mn_0^4 \,, \ 
k_i \neq (0,0,0,0)\ ,
\label{onbv}
\een
where the $\,\phi^\al_{(k_1,\ldots, k_v)}(p_1,\ldots, p_v)\,$ 
have been introduced in \eqref{nsbasis}.
}

\vspace{.2cm}
\noindent
    {\textit{Proof}}. We can use the presentation from (\eqref{mSs})
 for the function
\ben
\hms_{v,n-v}\,\bigl[\phi^\al_{(k_1,\ldots, k_v)}\,
\1^{n-v}\bigr]\ .
\label{hms}
\een
Due to the orthonormality
of the functions $\phi^\al_{(k_1,\ldots, k_v)}$
the scalar product 
of this function with itself in $\,{\cal H}^\al_n\,$
contains $\left( n \atop v \right)$ nonvanishing terms,
on expanding the double sum implied when applying
$\,\hms\,$ to the square brackets in the  states \eqref{hms}. 
Each of these terms takes the value $1$, since    
the functions 
$\,\phi^\al_{(k_1,\ldots, k_v)}(p_1,\ldots, p_v)\,$ 
and $\,\1^{n-v}(p_{n-v+1},\ldots,p_{n})\,$ 
are normalised to $1$ in the Hilbert spaces $\,{\cal S}^\al_{v,0}\,$ 
and $\,{\cal S}^\al_{0,n-v}\,$. This explains the normalisation factor
$\,\left( n \atop v \right)^{-\frac12}\,$. 
\hfill \proofSymbol

\vspace{.2cm}
\noindent
{\remark
As regards \textbf{notation}, we will make the following simplifications
for the rest of this section and in section \ref{sec:analpre}~:\\
1. We will mostly suppress the lower indices from 
$\,\hms_{v,n-v}\,$, and just write $\,\hms\,$,  
if these indices are clear from the context.\\
2. We will also  most often suppress the upper index $\al$ from the notation.
}

{\lemma
\label{decop} 
For 
\[
\psi_{n_i}^{(i)}\in {\cal S}_{n_i} \ , \quad 1 \le i \le r\ ,
\]
with the decompositions
\[
\psi_{n_i}^{(i)}\,=\,  \sum_{v_i=0}^{n_i} \ms \Bigl[\psi^{(i)}_{v_i} 
\1^{n_i-v_i}\Bigr]\ ,\quad \psi^{(i)}_{v_i} \in {\cal S}_{v_i,0}\ ,
\]
we have
\[
\psi_n \,:=\, 
\ms \Bigl[ \bigotimes_i \psi_{n_i}^{(i)} \Bigr] \,=\,
\sum_{v=0}^{n} \ms \Bigl[\psi_{v} \,
\1^{n-v}\, \Bigr]\ ,\quad n= \sum_i n_i\ ,
\]
with
\[
\psi_{v} \,=\, \sum_{\{ v_i\},\sum v_i =v} \ms \,
\Bigl[\, \bigotimes_i \psi^{(i)}_{v_i}  \,\Bigr]\ .
\]
}

\vspace{.2cm}
\noindent
{\textit{Proof}}.  
We find by linearity and from Lemma \ref{lms}
\[
\ms \Bigl[ \bigotimes_i \psi_{n_i}^{(i)}  \Bigr] \,=\,
\sum_{\{v_i\}} 
\ms \Bigl[\bigotimes_i \psi^{(i)}_{v_i} 
\1^{n_i-{v_i}}\Bigr] 
 \,=\,
\sum_{v=0}^{n}\ \, \sum_{\{ v_i\},\sum v_i =v}
\ms \Bigl[\, \Bigl(\bigotimes_i \psi^{(i)}_{v_i}\,\Bigr) \1^{n-v}
\Bigr] 
\]
\[
\,=\,\ms \biggl[\sum_{v=0}^{n}\ \,  \sum_{\{ v_i\},\sum v_i =v}
\ms \Bigl[\, \Bigl(\bigotimes_i \psi^{(i)}_{v_i}\,\Bigr) \1^{n-v}
\Bigr]\biggr] 
\,=\,
\sum_{v=0}^{n}\ \,\ms \biggl[ \biggl( \sum_{\{ v_i\},\sum v_i =v}
\ms  \Bigl(\bigotimes_i \psi^{(i)}_{v_i}\,\Bigr)
 \biggr) \1^{n-v}\biggr]
\]
\[
 \,=\,
\sum_{v=0}^{n} 
\ms \; \Bigl[\,\psi_{v} \,
\1^{n-v}
\, \Bigr]\ .
\]
\hfill \proofSymbol

\vspace{.2cm}


{\lemma 
For $\psi_n \,\in\, \cS_n\,$,  using the 
orthogonal decomposition \eqref{decos}, and
setting 
\[
\psi_n^{[n-v]}(p_1,\ldots, p_n)
\]
\vspace{-.4cm}
\ben
: = \ms\Bigl[ 
\Bigl(\!\!\int \psi_n(p_1,\ldots, p_v , p'_{v +1},\ldots, p'_n)
\, \d\mu(p'_{v+1},\ldots, p'_n)\Bigr) \1^{n-v}(p_{v+1},\ldots,p_n ) \Bigr]
\label{nnv}
\een
we have 
\ben
\psi_{v,n-v} = \left( n \atop v\right) \sum_{\mu=0}^v (-1)^{v-\mu} 
\left( v \atop \mu\right)\,  
\psi_n^{[n-\mu]}\ .
\label{deco}
\een
\label{zerl}
}

\vspace{.2cm}

\noindent
{\textit{Proof}}. 
To prove  \eqref{deco}, we use \eqref{decos}
on the r.h.s. of \eqref{nnv} which gives
\[
\psi_n^{[n-v]}
\]
\[
\,=\, 
 \ms\int\Bigl[\, \sum_{\mu=0}^v 
\Bigl(\ms\, [\,\psi_\mu \ \1^{n-\mu}\,]
(p_1,\ldots, p_v , p'_{v +1},\ldots, p'_n)
 \ \d\mu(p'_{v+1},\ldots, p'_n)\,\Bigr) \1^{n-v}(p_{v+1},\ldots,p_n ) 
\,\Bigr]\ \ \,
\]
\[
\,=\,
\sum_{\mu=0}^v \ms\,\Bigl[\, 
\int \Bigl(\ms\, [\,\psi_\mu  \ \1^{n-\mu}\,]
(p_1,\ldots, p_v , p'_{v +1},\ldots, p'_n)
 \ \d\mu(p'_{v+1},\ldots, p'_n)\Bigr) \1^{n-v}(p_{v+1},\ldots,p_n ) 
 \,\Bigr]
\ .
\]
Here we used orthogonality \eqref{ortho},
 to limit the sum to $\,\mu \le v\,$, 
and linearity
of $\,\ms\,$.
Using again \eqref{ortho}
 gives for $\,\mu \le v\,$
\[
\Bigl( \int \ms\,[\,\psi_\mu  \ \1^{n-\mu}\,]
(p_1,\ldots, p_v , p'_{v +1},\ldots, p'_n)
 \ d\mu(p'_{v+1},\ldots, p'_n)\Bigr)\,\1^{n-v}(p_{v+1},\ldots,p_n )
\]
\[
\,=\, 
\frac{1}{n!}\ \left( v \atop \mu\right) \ \mu! \ (n-\mu)! \
\ms\, [\,\psi_\mu  \ \1^{v-\mu}\,](p_1,\ldots, p_v)\Bigr]
\1^{n-v}(p_{v+1},\ldots,p_n )
\ .
\]
The prefactor $\,\frac{1}{n!}\, $ stems from the normalisation
of $\,\ms\,$. The integration only gives a nonvanishing result
(due to \eqref{ortho})
if $\,\psi_\mu\,$ carries arguments among $\,(p_1,\ldots,p_v)\,$.
There are $\,\left( v \atop \mu\right) \ \mu! \ (n-\mu)! \,$ permutations
of the $n$ arguments of $\,[\,\psi_\mu  \ \1^{v-\mu}\,]\,$ which realise
this case. The result is completely symmetric in 
$\,(p_1,\ldots,p_v)\,$ so that we were allowed to
replace the individual contributions
by $\,\ms\, \Bigl[\,\psi_\mu  \ \1^{v-\mu}\,](p_1,\ldots, p_v)\,$.
On applying $\,\ms\,$ again and summing over $\,\mu\,$ we  get
using Lemma \ref{lms}
\ben
\psi_n^{[n-v]}\,=\,\sum_{\mu =0}^v \
\frac{1}{n!}\ \left( v \atop \mu\right) \ \mu! \ (n-\mu)! \
\ms\, [\,\psi_\mu  \ \1^{n-\mu}\,]\,=\,
\frac{v!}{n!} \  
\sum_{\mu=0}^v \frac{(n-\mu)!}{(v-\mu)!}\
\psi_{\mu,n-\mu}\ .
\label{lisy2}
\een
The linear $\, n \times n\,$ system \eqref{lisy2} is invertible 
since 
the corresponding matrix
is triangular with positive constants $\,\left( n \atop v\right)^{-1}\,$
on the diagonal. So the solution is unique. 
We may find it by induction in $\,v\,$. It is simpler to
insert the r.h.s. of \eqref{deco} on the
r.h.s.  of \eqref{lisy2}. Then  we get
\[
\frac{v!}{n!} \  \sum_{\mu=0}^v \frac{(n-\mu)!}{(v-\mu)!}\
\left( n \atop \mu\right) \sum_{\la=0}^\mu (-1)^{\mu-\la} 
\left( \mu \atop \la\right)\,  
\psi_n^{[n-\la]}
\,=\,
\sum_{\la=0}^v \psi_n^{[n-\la]} 
\sum_{\mu=\la}^v (-1)^{\mu-\la} 
\left( \mu \atop \la\right)\,\left( v \atop \mu\right)\ .
\]
As a consequence of Euler's finite difference theorem,
see e.g. ~\cite{Gould}, (10.1) and (10.9),
we have
\[ 
\sum_{\mu=\la}^v (-1)^{\mu-\la} 
\left( \mu \atop \la\right)\,\left( v \atop \mu\right) = \de_{\la \nu}\ ,
\]
which proves that \eqref{deco} equals indeed the unique solution
of  \eqref{lisy2} and thus 
proves the  assertion.

\hfill \proofSymbol

\vspace{.2cm}

{\lemma
For $\psi_n \,\in \, \cS_n^{\infty}\,$ we have 
\ben
\psi_{n}^{[n-v]} \in \,\bigoplus_{\mu =0}^v  {\cal S}_{\mu,n-\mu}^{\infty}\ ,\ \  \
||\psi_{n}^{[n-v]}  ||_{\infty} \,\le\,  || \psi_{n} ||_{\infty}
\,\le \, \sum_v ||\psi_{v,n-v} ||_{\infty}\ ,  
\label{infnns}
\een
\ben
||\psi_{v,n-v} ||_{\infty} \, \le \, \left( n \atop v\right)
2^v\
\,||\psi_{n} ||_{\infty} \ ,  
\label{egauss}
\een
and for $\,\psi^{(i)}_{n}  \in {\cal S}_{n}^{\infty}\ ,\ \ 1 \le i \le r\,$, 
\ben
||\; \bigl[\prod_i \psi^{(i)}_{n}  \bigr]_{v,n-v}\, ||_{\infty} \, 
\le \, \left( n \atop v\right) 2^{v} 
\,\prod_i ||\, \psi^{(i)}_{n} \,||_{\infty} \ .
\label{cor1}
\een
\label{compbd}
}
\vspace{.2cm}

\noindent
{\textit{Proof}}.
All statements are obvious from
 \eqref{decos}, \eqref{nnv} and \eqref{deco}.
The last
 inequality follows from
\[
||\; \prod_i  \psi^{(i)}_{n} \,||_{\infty} \, 
\le \,\prod_i  ||\, \psi^{(i)}_{n} \,||_{\infty}\ .
\]
\hfill \proofSymbol

\subsection{Translation invariance and lifting}
\label{til}

Bose symmetry together with translation invariance 
lead to Schwinger n-point functions 
which are fully symmetric under permutations of 
their four-momentum arguments $\, p_1,\ldots, p_n\,$, 
where those are restricted to the ``hypersurface''
\ben
\mk^{4(n-1)} := \{(p_1, \ldots, p_n) \in \mr^{4n}\ |\  p_1 +\dots+p_n =0\,\}\ .
\label{complex}
\een
In our procedure of constructing such n-point functions from
the two-point function, we need a prescription of extending or
``lifting''
them (analytically) from $\,\mk^{4(n-1)} \,$ to $\,\mr^{4n}\,$.   
{\definition
A real function $\,\ti \psi_{n-1}\,$ on $\,\mk^{4(n-1)}\,$
 is in $\,{\cal S}(\mk^{4(n-1)})\,$ iff the following holds true\,:\\
Let for $ 1 \le i \le n\,$
\ben
\nonumber
\begin{split}
\,{\cal F}_i\  :\  \mr^{4(n-1)} \ \ &\longrightarrow\ \ \mk^{4(n-1)}\\
(p_1,\ldots,p_{n-1})\ \  &\mapsto \ \ 
(p_1,\ldots,p_{i-1}, -\sum_{j=1}^{n-1} p_j, p_i, \ldots, p_{n-1})
\end{split}
\een
and set 
\[
\psi_{n-1}^{(i)} :=
[\ti \psi_{n-1} \circ{\cal F}_i] \ .
\]
Then 
\ben
\,\psi_{n-1}^{(i)} \in {\cal S}_{n-1} \quad  
\mbox{ and } \quad 
\psi_{n-1}^{(i)}(p_1,\ldots,p_{n-1}) = \psi_{n-1}^{(j)} (p_1,\ldots,p_{n-1})
\quad \forall\  1\le i,\,j \le n\ .
\label{sk}
\een
Furthermore we set
$ \,{\cal S}^\infty(\mk^{4(n-1)})\,:=\,{\cal S}(\mk^{4(n-1)}) \cap
L^\infty(\mk^{4(n-1)})\ $.
\label{translin}
}

\vspace{.2cm}
{\definition
For $\, \ti \psi_{n-1}  \in {\cal S}^\infty(\mk^{4(n-1)})\,$ 
we define its \textbf{minimal lifting} $\,\psi_{n}(\ti\psi_{n-1})\,$ 
through
\ben
[\psi_{n}(\ti\psi_{n-1})](p_1,\ldots,  p_n) \,=\, 
\frac{1}{n} \sum_{i=1}^n \, [\ti \psi_{n-1}\circ{\cal F}_i] 
(p_1,\ldots,p_{i-1}, p_{i+1}, \ldots, p_n) \1(p_i)\ .
\label{lifts}
\een 
\label{dminlift}
}
  
\vspace{.2cm}
{\lemma
The minimal lifting  $\,\psi_{n}(\ti\psi_{n-1})\,$ 
satisfies
$\,\psi_{n}(\ti\psi_{n-1}) \in  {\cal S}^\infty_n(\mr^{4n})\,$ and 
\ben
\psi_{n}(\ti\psi_{n-1})\Bigr|_{\mk^{4(n-1)}}\,= \,\ti\psi_{n-1}\ , \quad
\, \psi_n(\ti\psi_{n-1})  
\in\bigoplus_{v =0}^{n-1}{\cal S}_{v,n-v} \,\subset {\cal S}_n\ .
\label{lift}
\een
Furthermore
\ben
 \ti\psi_{n-1}\,\in C^\infty(\mk^{4(n-1)})\ \;  \Rightarrow \ \; 
\psi_n(\ti\psi_{n-1})\,\in C^\infty(\mr^{4n})\ ,
\label{diff}
\een   
\ben
||\psi_n(\ti\psi_{n-1})||_\infty\,  =\,  ||\ti\psi_{n-1}||_\infty \ .
\label{liftprop}
\een 
\label{minlift}
}

\vspace{.2cm}

\noindent
{\textit{Proof}}. 
For $\si \in S_n\,$ we have 
\ben
\nonumber
\begin{split}
[\psi_{n}(\ti\psi_{n-1})](p_{\si(1)},\ldots,  p_{\si(n)}) 
&\,=\, 
\frac{1}{n} \sum_{i=1}^n \, [\ti \psi_{n-1}\circ{\cal F}_i] 
(p_{\si(1)},\ldots,p_{\si(i-1)}, p_{\si(i+1)}, \ldots, p_{\si(n)}) \1(p_{\si(i)})\\
&=\,
\frac{1}{n} \sum_{i=1}^n \, [\ti \psi_{n-1}\circ{\cal F}_{\si(i)}] 
(p_{1},\ldots,p_{\si(i)-1}, p_{\si(i)+1}, \ldots, p_{n}) \1(p_{\si(i)})
\\
&=\,
\frac{1}{n} \sum_{i=1}^n \, [\ti \psi_{n-1}\circ{\cal F}_i] 
(p_{1},\ldots,p_{i-1}, p_{i+1}, \ldots, p_{n}) \1(p_{i})\\
&\,=\,
[\psi_{n}(\ti\psi_{n-1})](p_{1},\ldots,  p_{n})\  ,
\end{split}
\een
where the second equality follows from \eqref{sk} and the third
is by rearrangement of the sum and by symmetry of the
$\,\psi_{n-1}^{(i)}\,$. \eqref{lift} follows from Definition
\ref{translin}, noting that all entries in the sum for 
$\psi_{n}(\ti\psi_{n-1})$ are orthogonal to ${\cal S}_{n,0}\,$.
\eqref{diff} follows from the definition of $\psi_n$. 
It is obvious from the definition that 
$\, ||\psi_n(\ti\psi_{n-1})||_\infty\,  \le\,  ||\ti\psi_{n-1}||_\infty\,$. 
On the other hand
\[
||\psi_n(\ti\psi_{n-1})||_\infty\, \ge\,
||\, \Bigl[\psi_n(\ti\psi_{n-1})\bigr|_{\mk^{4(n-1)}}\Bigr]\,||_\infty\,=\,
||\ti\psi_{n-1}||_\infty \ ,
\]
due to \eqref{lift}. This implies \eqref{liftprop}. 
\hfill  \proofSymbol

\vspace{.2cm}
\noindent
\noindent
    {\remark \label{rm3}  We call our lifting procedure \textbf{minimal}
      because it is designed so as not to introduce any new
      momentum dependence in the functions
      $\,\psi_n(\ti\psi_{n-1})\,$. This corresponds to a
      principle of  \textbf{maximal locality} for the construction
      of the lifted functions. E.g. functions of the form
\[
\psi_n(\ti\psi_{n-1}) + (p_1 +\ldots +p_n)^2 f_n
\] 
with $\,f_n\in {\cal S}_{n}\, $, are also in 
$\,{\cal S}_n\,$ and satisfy 
\[
[\psi_n(\ti\psi_{n-1}) 
+ (p_1 +\ldots +p_n)^2 f_n\,]\Bigl|_{\mk^{4(n-1)}}
\,= \,\ti\psi_{n-1}\ .
\]
Their derivatives may not vanish even if those of 
 $\,\psi_n(\ti\psi_{n-1})\,$ do.    } 
\vspace{.2cm}

{\lemma
  Let $\psi_{n_i}\,\in\, {\cal S}(\mk^{4(n_i-1)})\,$, $ i =1,2\,$.
  Define
  \[
  \psi_n(p_1,\ldots,p_{n})\,:=\,
  \ms[\psi_{n_1}(p_1,\ldots,p_{n_1-1},p' )\,
    \psi_{n_2}(-p', p_{n_1},  \ldots, p_{n})\,]\Bigr|_{\mk^{4(n-1)}}\ ,
  \ \ n_1+n_2=n+2\ .
  \]
  Here the symmetrisation is w.r.t. the independent momenta
  $p_1,\ldots,p_{n_1-1},p_{n_1}, \ldots,p_n\,$. Then we have
    \[
  \psi_n\,\in \,{\cal S}(\mk^{4(n-1)})\ .
   \]
  \label{k1k2}
    }
\vspace{.2cm}

\noindent
{\textit{Proof}}.
The functions $\,\psi_{n_1}(p_1,\ldots,p_{n_1-1},p' )\,$
and $\, \psi_{n_2}(-p', p_{n_1},  \ldots, p_{n})\,$
can be parametrised on $\,\mk^{4(n_1-1)}\,$ resp. on
$\,\mk^{4(n_2-1)}\,$ as functions of
$\,(p_1,\ldots,p_{n_1-1})\,$ and $\,(p_{n_1}, \ldots, p_n)\,$,
and they are fully symmetric
w.r.t. these four-momenta due to Definition \ref{translin}. For
\[
(p_1,\ldots,p_{n_1-1},p' )\in \mk^{4(n_1-1)}\ , \quad
  (-p', p_{n_1},  \ldots, p_{n})\in \mk^{4(n_2-1)}
\]
we have
$(p_1,\ldots,p_{n_1-1}, p_{n_1},  \ldots, p_{n})\in \mk^{4(n-1)}\,$,
and the function 
$\,\psi_n(p_1,\ldots,p_{n})\,$ on
$\,\mk^{4(n-1)}\,$ is then fully symmetric w.r.t.
its arguments due to symmetrisation.
\hfill  \proofSymbol


\section{Preimages preserving analyticity}
\label{sec:analpre}

 \vspace{.2cm}

In order to construct solutions
of the flow equations \eqref{cagfe}, which are  analytic
w.r.t. the flow parameter and  the four-momentum variables,
we will use the same technique as in ~\cite{Kopper},
namely we construct the functions $\,\ell_{n+2}\,$
as  functions of the  $\,\pa^w_\mu \ell_{n'}\,,\ n' \le n\,$,
starting from a given regular two-point function
$\,\ell_{2}(\mu;q_1,q_2)\,$,
 such that the analyticity
 and certain boundedness properties are  preserved.
   Since  $\,\ell_{n+2}\,$ appears 
   in integrated form in the FE \eqref{cagfew}
   for  $\,\frac{d}{d\mu} \ell_n\,$,
 this construction is not unique.
Imposing sufficiently strong  regularity
requirements for the solutions, uniqueness for given boundary conditions
at $\mu =0\,$ can however be established, using similar arguments as were
used in ~\cite{KW} in the mean field approximation. 
Namely when requiring analyticity,
different choices for the construction of $\,\ell_{n+2}\,$ will
lead to different boundary conditions and then to different
solutions. This concerns the lifting procedure considered in the
previous section and also the construction of the integral presented in
this section.
Regularity properties of the solutions of the type considered here
are in any case required for a quantum field theory.
 For a precise formulation of our uniqueness
statement see Theorem \ref{thm1} in section \ref{smgs}.

We now proceed by defining mappings corresponding
to the integration operation acting on the l.h.s. of the FEs 
\eqref{fescale} or \eqref{cagfe}.
The aim is to construct, for any given function
in a suitable class supposed to represent the r.h.s. of
\eqref{fescale} or \eqref{cagfe}or \eqref{cagfew},
a function $\,\ell_{n+2}\,$ which  is a mapped onto
this given function by the integration operation. 

{\definition 
The family $\,\{\C_{n+2}\}_{n \ge 0}\,$ of graded linear mappings of degree $-2$ 
on the vector  spaces $\,\S_{n+2}^\infty\,$ is defined as
follows:
\[
\C_{n+2}\, :\,  {\cal S}^\infty_{n+2}\, \rightarrow {\cal S}^\infty_{n}\quad 
\mbox{for } \ \ n \in \mathbb{N}_0 \ ,
\]
\ben
[{\cal C}_{n+2}(\psi_{n+2})](q_1,\ldots,q_n)\,:=\,\int_{q} \dot{\C}(q)\,
\psi_{n+2}(q_1,\ldots,q_n, q,-q)\  ,
\quad \dot {\cal C}(q) := \dot {\cal C}^1(q)
\label{C}
\een
remembering \eqref{norm}. For ease of notation we set
\ben
\C_n = 0 \quad \mbox{for}\quad  n < 2\ .
\label{kl}
\een 
For arbitrary index $\al>\alo >0\,$ we set correspondingly 
\ben
    [{\cal C}^{(\al)}_{n+2}(\psi_{n+2})](p_1,\ldots,p_n)
    \,:=\,\int_{q} \dot{\C}^\al(p)\,
\psi_{n+2}(p_1,\ldots,p_n, p,-p)\  .
\label{Cal}
\een
\label{grading}
}

\noindent
From \eqref{norm} we have
\ben
{\cal C}^{(\al)}_{n+2}(\1^{n+2}) \,=\, \1^{n} \ ,\ \ \mbox{ and } \ \
| | \,{\cal C}^{(\al)}_{n+2}(\psi_{n+2})\,| |_{\infty}\,\le \, 
||\,\psi_{n+2}\,| |_{\infty}  
\label{infty}
\een
so that $\C^{(\al)}_{n+2}$ is bounded by $1$ w.r.t. the $\,L^\infty$-norm.\\[.1cm]
Note however that $\, \C^{(\al)}_{n+2}\,$ is unbounded w.r.t. the $\,L^2$-norm. 
For $\,n+2=4$ this can be verified considering the example
$\, \psi_4(p_1,p_2,p_3,p_4) 
\,=\,  
\mathbb{S}\Bigl[ \,[(p_1+p_2)^2 +\vep^2]^{-1/2}\,1 \!\!\! 1(p_3,p_4)\Bigr]\,$
for $\,\vep\,$ arbitrarily small.

\vspace{.2cm}

\noindent
{\remark
For ease of \textbf{notation}
we will suppress the indices $n$ in the notation of 
the mapping $\,\C_{n}\,$
 unless this could give rise to confusion.  So we will write 
$\, \C \ \mbox{ instead of }\  \C_{n} \,$ and 
$\, \C^\ka \,$ instead of  $\,\C_{n-2\ka+2}
\ldots  \circ \C_{n-2}\circ  \C_{n} 
\;$.
Apart from a few comments at the end of this section,
we will also set  $\,\al=1\,$ subsequently so that the upper index
in $\,{\cal C}^{(\al)}_{n}\,$ does not appear. 
With hindsight to the transformation \eqref{scaling2}
we will generally denote the momenta $q_i$ instead of
$p_i$ if we have chosen $\,\al=1\,$.
}

\vspace{.2cm}
\noindent
Let $\, \psi_{v,n-v } \in {\cal S}_{v,n-v }^\infty\,$. 
For the rest of this section we will always write
\ben
\psi_{v,n-v}(q_1,\ldots,q_{n})\,=\, 
\hms\Bigl[ 
\chi_{v} (q_1,\ldots,q_{v })\, 
\1^{n-v } (q_{v +1},\ldots,q_{n})\,\Bigr]\ .
\label{psinnu}
\een
Comparing to \eqref{decos} (for $\al =1\,$) and remembering
the definition \eqref{hatms}, this means
that $\,\chi_{v} \in \S^\infty_{v,0}\,$
is normalised so as to satisfy  
\ben
\psi_{v} (q_1,\ldots,q_{v })\,=\, \left( n \atop v \right)\, 
\chi_{v} (q_1,\ldots,q_{v })\ . 
\label{norm3}
\een

{\lemma 
Let $\,n \ge 2\,$. For   $\, \psi_{0 ,n }\,\in {\cal S}_{0,n}^\infty\ 
\Leftrightarrow \ \psi_{0 ,n }\,=\,\chi_0 \,\1^n\,$
with $\,\chi_0 \in \mr\,$, 
we have 
\ben
\C(\psi_{0 ,n })\,=\, \chi_0\, \C(\1^n)\,=\, \chi_0\, \1^{n-2}\ .
\label{0}
\een
For  $\, \psi_{1 ,n-1 } \in {\cal S}_{1 ,n-1 }^\infty\,$ holds
\ben
\C(\psi_{1 ,n-1 })\,=\, 0\ .
\label{1}
\een
For $\, \psi_{n,0}\,\in {\cal S}_{n,0}^\infty\,$ we have
\ben 
\C(\psi_{n,0}) \in {\cal S}_{n-2,0}^\infty\ .
\label{3}
\een 
For $\, 0 \le v \le n\,$ we obtain generally
\ben
\C(\psi_{v,n-v})\,=\,  
\hms\Bigl[\,
\chi_{v}\,\C(\1^{n-v} )\,\Bigr] \,+\,
\hms\Bigl[ \,\C(\chi_{v})\, \1^{n-v } \,\Bigr]\ .
\label{gg}
\een
\label{c1}
}

\vspace{.2cm}

\noindent
\textit{Proof}. 
The relations  
\eqref{0} and \eqref{3} follow directly form the definition of $\,\C\,$,
\eqref{1} is a consequence of orthogonality \eqref{ortho}.\\ 
To prove \eqref{gg} we write
\ben
\C(\psi_{v,n-v})\,=\, \int_q   \dot{\C}(q)\,[\,\hms
\,\bigl(\chi_{v} \,  \1^{n-v }) \,\bigr](q_1,\ldots,q_{n-2},q,-q)\ .
\een
Among the 
\[ 
\left({ n \atop \nu}\right)
\]
shuffles  implied by the action of
$\,\hms\,$, there are 
\ben
\left({ n-2 \atop n-\nu-2}\right)\,=\, 
\left({ n \atop \nu}\right)\ \frac{(n-\nu)(n- \nu-1)}{n (n-1)}
\label{t2}
\een
in which the arguments $q,\,-q\,$ are both attributed to $\,\1^{n-\nu}\,$,
and  
\ben
\left({ n-2 \atop \nu-2}\right)\,=\, 
\left({ n \atop \nu}\right)\ \frac{\nu(\nu-1)}{n (n-1)}\ ,
\label{t1}
\een
in which they are both attributed to $\,\chi_{v} \,$,
and finally
\ben
2\, \left({ n-2 \atop \nu-1}\right)\,=\, 
2\, \left({ n \atop \nu}\right)\ \frac{\nu(n-\nu)}{n (n-1)}
\label{t3}
\een
in which $\,q\,$ is attributed to $\,\1^{n-\nu}\,$
and  $\,-q\,$  is attributed to $\,\chi_{v} \,$, 
or vice versa.
Consistently we find 
\[
\left({ n \atop \nu}\right)\,=\,
\left({ n-2 \atop \nu-2}\right)\,+\,
\left({ n-2 \atop n-\nu-2}\right)\,+\,
2\, \left({ n-2 \atop \nu-1}\right)\ .
\]
Attributions of the type related to \eqref{t3}, 
give $0$ on application of $\C\,$, due to \eqref{ortho}.
Remembering \eqref{hatms} we therefore get \eqref{gg}
 \[
\C(\psi_{v,n-v})\,=\, 
\hms\Bigl[\chi_{v}\, \C( \1^{n-v }) \,\Bigr]+\,
\,  \hms\Bigl[\C(\chi_{v})\, \1^{n-v }\Bigr]\ .
\]
 \hfill  \proofSymbol
 
\vspace{.2cm}


\noindent
As a consequence of Lemma \ref{c1} we can now give an explicit 
expression for a
preimage of a given $\,\psi_{v,n-v } \in {\cal S}_{v,n-v  }^\infty\,$
under $\,\C\,$.

{\prop
For given $\; \psi_{v,n-v }\,=\,  \hms\Bigl[\chi_{v}\, \1^{n-v}\Bigr]\,  
\in {\cal S}_{v,n-v}^\infty\,$ as in \eqref{psinnu},
we define 
\ben
\Phi_{n+2}(\psi_{v,n-v}) \,:=\, \sum_{\rho =0}^{[\frac{v }{2}]} (-1)^\rho \,
\hms\Bigl[\, \C^\rho (\chi_{v})\,  
\, \1^{n+2-v +2\rho}\Bigr] \  . 
\label{phipsi}
\een
Then $\,\Phi_{n+2}(\psi_{v,n-v})\in {\cal S}_{n+2}^\infty\,$, and 
\ben
\C(\Phi_{n+2}(\psi_{v,n-v }))\,=\, \psi_{v,n-v}\ .
\label{urbildgl}
\een
\label{urbild}
}
\noindent
\textit{Proof}. 
It is straightforward to verify that 
$\,\Phi_{n+2}(\psi_{v,n-v}) \in \S_{n+2}^{\infty}\,$
since  $\,\psi_{v,n-v} \in \S_{n}^{\infty}\,$.\\[.1cm]
We  evaluate $\,\C(\Phi_{n+2}(\psi_{v,n-v}))\,$  using  Lemma \ref{c1}.
\[
\C(\Phi_{n+2}(\psi_{v,n-v}))\,=\, 
\hms\Bigl[ \chi_{v}  \ \C(\1^{n+2-v })\Bigr]\ +\ 
\hms\Bigl[\C( \chi_{v}) \ \1^{n+2-v}\Bigr] 
\]
\[
+ \,\sum_{\rho =1}^{[\frac{v }{2}]} (-1)^\rho \,
\Biggl(
\hms\Bigl[ \C^{\rho}(\chi_{v})  \,\,\C(\1^{n+2-v +2\rho})\Bigr]
+ \hms\Bigl[ \C( (\C^{\rho}(\chi_{v}))  \,\1^{n+2-v +2\rho})\Bigr]\Biggr)
\] 
\[
=\,
\hms\Bigl[ \chi_{v}  \ \1^{n-v}\Bigr]
\]
\[
+\ \hms\Bigl[\C( \chi_{v}) \ \1^{n+2-v }\Bigr]
+\,
\sum_{\rho =1}^{[\frac{v }{2}]}  (-1)^\rho \, 
\Biggl(
\hms\Bigl[\C^{\rho} (\chi_{v})  \,\,  \1^{n-v+2\rho}\Bigr]
 \ +\
\hms\Bigl[ \C^{\rho+1} (\chi_{v})  \,\, \1^{n+2-v +2\rho}\Bigr] \Biggr)\ .
\]
The sum over $\,\rho\,$
telescopes, and using \eqref{kl} which implies
\[
\C^{[\frac{v }{2}]+1} (\chi_{v}) = 0\ ,
\]
we get
\[
\C(\Phi_{n+2}(\psi_{v,n-v}))
\,=\, \hms \Bigl[\chi_{v}\, \1^{n-v}\Bigr]\,=\, \psi_{v,n-v}\ .
\]
 \hfill  \proofSymbol

\vspace{.3cm}
\noindent
On summing over $\,v\,$ we  get directly
{\prop
For  $\, \psi_n\,\in \,{\cal S}_{n}^\infty\,$, 
with the orthogonal decomposition
$\,\psi_{n}\,=\,\sum_{v=0}^n \psi_{v,n-v}\, ,$\\
$\,\psi_{v,n-v}\,=\, \hms\Bigl[\chi_{v}\, \1^{n-v}\Bigr]\,$,
we set
\ben
\Phi_{n+2}(\psi_n) \,:=\,\sum_{\nu=0}^n \sum_{\rho =0}^{[\frac{v }{2}]} (-1)^\rho \,
 \hms\Bigl[\, \C^\rho (\ \chi_{v})\,  
\, \1^{n-v +2\rho+2}\Bigr]  \ .
\label{phipsi1}
\een
Then
\ben
\C(\Phi_{n+2}(\psi_n) )\,=\, \psi_n \quad \mbox{and}\quad 
\Phi_{n+2}(\psi_n) \,\in \,{\cal S}_{n+2}^\infty\ .
\label{urbildgl2}
\een
\label{urbild2}
}
{\definition
For a function $\psi_n \in \,{\cal S}^\infty_{n}\,$ we call
$\,\Phi_{n+2}(\psi_n)\,$ as defined in \eqref{phipsi1},
the \textbf{minimal extension}
of  $\,\psi_n\,$ to $\,{\cal S}^\infty_{n+2}\,$.
Again minimal extension corresponds to 
 \textbf{maximal locality} in the sense of remark \ref{rm3}
in the choice of the extension. 
\label{minext}
}

\vspace{.1cm}

\noindent
{\prop
For  $\, \psi_n\,$ as in  Proposition \ref{urbild2},
decomposing  $\,\Phi_{n+2}(\psi_n)\,$ into orthogonal components
\ben
\Phi_{n+2}(\psi_n) = \sum_{\nu=0}^{n+2} [\Phi_{n+2}(\psi_n)]_{v,n+2-v}\ \, ,
\quad  [\Phi_{n+2}(\psi_n)]_{v,n+2-v}\,\in\,
{\cal S}_{v,n+2-v}^\infty\ , 
\label{urbildv}
\een
we have the bounds
\ben
\!
||[\Phi_{n+2}(\psi_n)]_{v,n+2-v}||_\infty \le \frac{n\!-\!v\!+\!1}{2}\!
\sup_{\ka \in \{v, v+2, \ldots, n^*(v)\}}\! 
\Biggl(\frac{\left( { n+2 \atop v}\right)}{\left( { n \atop \ka} \right)} 
||\,\ms\Bigl[\, \C^{\frac{\ka-v}{2}}\,(\psi_{\ka})\1^{n+2-v}]\,||_\infty\Biggr)
\ , 
\label{!!}
\een
\label{bdphinu}
where we set $\,n^*(v) = v +2[\frac{n-v}{2}]\,$. 
}
\vspace{.25cm}

\noindent
\textit{Proof}. 
Proceeding as in the proof of 
Proposition \ref{urbild} we have 
\ben
\begin{split}
\qquad\  [\Phi_{n+2}(\psi_n)]_{v,n+2-v} &\,=\,
\sum_{\ka=0}^n \sum_{\rho =0}^{[\frac{\ka }{2}]} (-1)^\rho \,
 \hms\Bigl[\, \C^\rho (\chi_{\ka})\,  
\, \1^{n-\ka +2\rho+2}\,
  \Bigr]\, \de_{\ka -2\rho, v}\\
&\,=\,\sum_{\ka = v, v+2, \ldots, n^*}
(-1)^{\frac12\, (\ka-v)} \,
 \hms\Bigl[\, \C^{\frac12\, (\ka-v)}\,(\chi_{\ka})\,  
\, \1^{n+2-v}  \Bigr]\\
&\,=\,
 \hms\,\biggl[\,\biggl(\, \sum_{\ka = v, v+2, \ldots, n^*}
(-1)^{\frac12\, (\ka-v)} \,
\C^{\frac12\, (\ka-v)}\,(\chi_{\ka})\,\biggr) \, \1^{n+2-v}\, \biggr]
\ .
\end{split}
\label{urbildcomp}
\een
Therefore we obtain the bound
\[
||\,[\Phi_{n+2}(\psi_n)]_{v,n+2-v}||_\infty\,\le\, \frac{(n-v+1)}{2}\
\sup_{\ka \in \{v, v+2, \ldots, n^*\}} 
||\,\hms\Bigl[\, \C^{\frac12\, (\ka-v)}\,(\chi_{\ka}) \1^{n+2-v}]\,||_\infty
\]
\ben
\,\le\, \frac{(n-v+1)}{2}\
\sup_{\ka \in \{v, v+2, \ldots, n^*\}}\, 
\Biggl(\frac{\left( { n+2 \atop v}\right)}{\left( { n \atop \ka} \right)} \
||\,\ms\Bigl[\, \C^{\frac12\, (\ka-v)}\,(\psi_{\ka})\1^{n+2-v}]\,||_\infty\Biggr)
\ ,
\label{!!!}
\een
where we used \eqref{hatms} and \eqref{norm3}.
\hfill  \proofSymbol

{\lemma 
The map
\[
\Phi_{n+2}\, : {\cal S}^\infty_{n}\ \rightarrow \ {\cal S}^\infty_{n+2}\ 
\]
is continuouus w.r.t. the $\,\sup$-norm.\\
If $\,\psi_n\,$ is locally analytic w.r.t. $\mu$ and to the momenta
$q_1, \ldots, q_n\,$, then $\,\Phi_{n+2}\,$ has the same properties
w.r.t. $\mu$ and to the momenta
$q_1, \ldots, q_n, q_{n+1},q_{n+2}\,$.  
\label{stetig}
}

\vspace{.2cm}
\noindent
\textit{Proof}. The statement follows directly from \eqref{phipsi1}
and from $\,|| \C ||_\infty \,\le\, 1\ .$\\
$\,\psi_n\,$ is uniformly bounded and since  $\,\Phi_{n+2}\,$
is obtained from  $\,\psi_n\,$ by sums
of functions, which are obtained from  $\,\psi_n\,$  by gaussian
integration and multiplication by unit functions, the
analyticity statements follow trivially.
\hfill  \proofSymbol

\vspace{.2cm}

 \noindent
Again we note that the mapping is not continuous w.r.t. the  $L^2$ norm.

\vspace{.2cm}
\noindent
The previous statements hold for any value of $\al >0\,$.
For $\, \psi_n\,\in \,{\cal S}_{n}^{\al,\infty}\,$ we can define
analogously the orthogonal decomposition
\ben
\psi^{\al}_{n}\,=\,\sum_{v=0}^n \psi^{\al}_{v,n-v}\ ,\quad
\psi^{\al}_{v,n-v}\,=\, \hms\Bigl[\chi^{\al}_{v}\, \1^{n-v}\Bigr] \ .
\label{alzerl}
\een
We then set
\ben
\Phi^{(\al)}_{n+2}(\psi_n)
\,:=\,\sum_{\nu=0}^n \sum_{\rho =0}^{[\frac{v }{2}]} (-1)^\rho \,
 \hms\Bigl[\, [\C^{(\al)}]^\rho (\chi^{\al}_{v})\,  
\, \1^{n-v +2\rho+2}\Bigr]  
\label{phipsial}
\een
and find as in the case $\al=1$
\ben
\C^{(\al)}(\Phi^{(\al)}_{n+2}(\psi^{\al}_n) )\,=\, \psi^{\al}_n
\quad \mbox{and}\quad 
\Phi^{(\al)}_{n+2}(\psi^{\al}_n) \,\in \,{\cal S}_{n+2}^\infty\ .
\label{urbildal}
\een

\section{Inductive bounds on the r.h.s of the FEs}
\label{bounds}
In this section we bound  inductively the l.h.s. of the FEs
\eqref{cagfe} or \eqref{cagfew}.
We use the decomposition into components $(v,n-v)$ 
according to Propositions \ref{urbild} and  \ref{urbild2},
noting that the integration operation appearing on the l.h.s.
of the FEs corresponds to the action of the operator $\C\,$ from  
\eqref{C}. The construction requires to know
these components in order to obtain the (components of the)
$(n+2)$-point-function. To perform the construction
we lift the r.h.s. of the FEs
from  a function $\ti \psi \in {\cal S}(\mk^{4(n-1)})\, $ to 
a function $\,\psi_n\in {\cal S}_{n}\,$ using
Lemma \ref{minlift}. Then we  construct $\,\Phi_{n+2}(\psi_n)\,$.
This function can be restricted to  $\,{\cal S}(\mk^{4(n+1)})\, $
and inserted on the r.h.s. of the FE in the next induction step.

The bounds 
which we  prove are strong enough to show the existence
of global nontrivial solutions for the $n$-point functions,
and to verify the Osterwalder-Schrader growth condition
~\cite{OS}. This condition is part of the sufficient conditions
formulated by Osterwalder and Schrader, which guarantee that a
Euclidean quantum field theory gives rise to corresponding Wightman
theory. It is called $E0'$ or $E0''$ in ~\cite{OS}.  Translated
to our context it implies that the $n$-point CAS should not grow
more rapidly as functions of $n\,$
than some power of $n!\,$. We will show that
this condition holds for our construction, the power can be
chosen to be equal to 1.
These conditions however are not strong enough to prove the existence
of the effective
action and its local analyticity w.r.t. the field variable.
As said before we leave the analysis of
these problems and related ones to the future.

{\remark
Note that due to Lemma \ref{minlift}, \eqref{liftprop}, 
we need not distinguish between the $L^\infty$-norms of
the functions defined on $\mk^{4(n-1)}\,$ and of their minimally
lifted versions on
$\mr^{4n}\,$. We will comment on this distinction only when necessary.
}

\vspace{.2cm}
\noindent
For the purpose of this section, 
assuming the notations from \eqref{decos}, \eqref{psinnu},
\eqref{norm3}, we write the
decomposition of the n-point functions and
their derivatives
$\, \pa^{w}\,\ell_n \,$ - remembering
the notations \eqref{w1} and \eqref{w2} -
into orthogonal components
appearing in \eqref{cagfe}, \eqref{cagfew} 
\ben
\pa^w \ell_n(\mu; q_{1},\ldots,q_{n}) \,=\, 
\sum_{v=0}^n [\pa^w \ell_n]_{v,n-v}(\mu; q_{1},\ldots,q_{n})
\ \ \mbox{ with } \ \
[\pa^w\ell_n]_{v,n-v}\,\equiv\,
\hms [\,\pa^w\chi_{v}^{(n)} \1^{n-v}\,]\ .
\label{lnw}
\een
We shall bound the functions\footnote{In this section we use
for clarity the notation $\,[\pa^w \ell_n]_{v,n-v}\,$ instead
of $\,[\pa^w \ell]_{v,n-v}\,$.} 
$ \,[\pa^w \ell_n]_{v,n-v}(\mu; q_{1},\ldots,q_{n}) \,$
inductively w.r.t. $\,n\,$.
{\remark
  Attention has to be paid to the fact that the components of the
    orthogonal decomposition $\, \pa^{w}\,\ell\,$
    are generally not equal to the corresponding derivatives
    of the  components of the
    orthogonal decomposition $\,\ell_n\,$, i.e. generally
    \[
    [\pa^w \,\ell_n\,]_{v,n-v}
    \,\neq\,\pa^{w}\, ([\,\ell_n\,]_{v,n-v})\ .
    \]
    The decomposition into the orthogonal sectors
    ${\cal S}_{v,n-v}\,$ does not commute with taking derivatives.
        }

{\prop
  We assume that the minimally lifted (see Definition \ref{dminlift})
  two-point function 
  $\,\ell_{2}(\mu; q_{1},q_{2})$ is uniformly (see remark \ref{analyt})
  locally analytic
in its arguments, and that the components
$\,[\pa^w \ell_2]_{v,2-v}\,$
of its  derivatives are bounded
\ben
\Bigl|\Bigl|\; \,[\pa^w \ell_2]_{v,2-v}\; 
\Bigl|\Bigl|_\infty\, \le 
K^{w+v}\,\frac{|w|!}{2^3 (|w|+1)^2\,}\ , \quad 0 \le v \le 2
\label{2ptbd}
\een
for   $\,K> 0\,$  sufficiently large
and for  $\,w \in \mn^{2+1}_0\,$.
Furthermore we assume  for given $\,n\in 2\mn\,$,
$\,w \in \mn^{n'+1}_0\,$, $\,4\le n' \le n\,$
\ben
\Bigl|\Bigl|\, \, [\pa^w \ell_{n'}]_{v,n'-v} \,\Bigr|\Bigr|_\infty\, \le 
K^{n'+v-2+|w|}\,\left( {n' \atop v} \right)\ 
\frac{(n'-2+|w|)!}{n'^3\, (|w|+1)^2\,}\ , \ \ 
0 \le v \le n'\ \,.
\label{nptbd}
\een
We call the r.h.s. of the  FE \eqref{cagfew}
$\,\pa^{w}\,\psi_n$. If the functions $\,[\pa^w \ell]_{v,n'-v}\,$ are in
$\,\S_{n'}\,$ 
and are uniformly
locally analytic w.r.t. their arguments $\,\mu, q_1,\ldots,q_{n'}\,$,
symmetric w.r.t. $\,q_1,\ldots,q_{n'}\,$
and invariant under the Euclidean group,
then so are the functions  $\,[\pa^{w}\,\psi_n]_{v,n-v}\,$. Furthermore
$\,[\pa^{w}\,\psi_n]_{v,n-v} \in \S_n\,$, and the functions
$\,[\pa^{w}\,\psi_n]_{v,n-v}\,$ satisfy the bounds
\ben
||\,\,[\pa^{w}\,\psi_n]_{v,n-v}\,||_\infty \le \, 
\frac12\, \left[\frac{1}{n^3}\  \left( {n+2 \atop v} \right)\,
\,+\,
\frac{1}{n^2}\,  \left( {n \atop v} \right)\,\right]
K^{n+v+|w|}\,\frac{(n+|w|)!}{(n+2)^3\,(|w|+1)^2}\  .
\label{rhsl}
\een
\label{prop4}
}
\vspace{.02cm}

{\remark
  Regarding the regularity assumptions
  for the two-point function
  we assume that for any real point $(\mu,q_{_{11}},\ldots,q_{_{14}},
  q_{_{21}},\ldots,q_{_{24}})$ with $\mu \ge 0\,$, 
  there exists an $\vep$-neighbourhood,
  with $\vep$ independent of the point chosen, such that the function
  $\ell_2$ is analytic in this neighbourhood. We do not presently see the
  physical motivation to deal with  weaker assumptions, which nevertheless seem
  to be implementable without much change. By induction we will realise
  straightforwardly   that the analyticity statements on the two-point
  function then are
  inherited, in the same sense as regards uniformity, by the $n$-point
  functions. \label{analyt}}

\vspace{.2cm}
\noindent
    {\textit{ Proof}}.
    The analyticity statements and Euclidean invariance
    are obvious from the fact that
    locally analytic functions Euclidean invariant functions
    form a ring, and from the fact
    that these properties hold for  the function $\, \dot C^1(q)\,$.
    Bose symmetry follows from the fact that the terms on the r.h.s. of
    FEs are either already symmetric or explicitly symmetrised.

    \noindent 
    To  bound the component $[\quad]_{v,n-v}\,$ of the
    quadratic part, i.e. the second line
    on the r.h.s. of \eqref{cagfew}
\ben
\!\! \frac{1}{n+1} \!\!\!\!
\sum_{n_{1}+n_{2}=n+2\atop w_1+w_2+w_3=w} \!\!\!\! c_{w_1,w_2,w_3}\,
\left[\mathbb{S} \Bigl( \pa^{w_1} \ell_{n_{1}}(\mu;q_{1},\ldots, q')\, 
\pa^{w_3}\dot{C}^{1}(q')\, 
 \pa^{w_2}\ell_{n_{2}}(\mu;-q',\ldots,q_{n})\Bigr) 
\right]_{v,n-v}\ ,
\label{quad}
\een
where $\,q'= - \,(q_{1} - \ldots - q_{n_1-1})\,$,
we proceed as follows:
We first bound  by  induction the product of the first two terms
within the symmetrised product, which constitute a
symmetric function of $\,q_{1}, \ldots, q_{n_1-1}\,$
\ben
\begin{split}
&\qquad \biggl|\,[\pa^{w_1} \ell_{n_{1}}](\mu;q_{1},\ldots, q')\
   [\pa^{w_3} \dot{C}^{1}](q') \Bigr]_{v_1,n_1-v_1}\,\biggr|\\
   &   \le  \sum_{0\le  v' \le v_1}
  \Bigl|\, [\pa^{w_1} \ell_{n_{1}}(\mu;q_{1},\ldots, q')]_{v',n_1-v'}\
   [\pa^{w_3}\dot{C}^{1}](q')]_{v_1-v',n_1-v_1+v'}\,\Bigr|\\
     &\le \sum_{0\le  v' \le v_1}K^{n_1+v'-2+|w_1|}\,
     \left( {n_1 \atop v'} \right) \,
     \frac{(n_1-2+|w_1|)!}{n_1^3 (|w_1|+1)^2}
     \left( {n_1 \atop v_1-v'}\right) \,
       2^{v_1-v'} \ \frac{K^{|w_3|}\ |w_3|!}{ (|w_3|+1)^2}\\
       &\le\,\sum_{0\le  v' \le v_1}
       K^{n_1+v_1-2+|w_1|+|w_3|}\,
       \left( {n_1 \atop v'} \right)  \left( {n_1 \atop v_1-v'} \right)
      \,
     \frac{(n_1-2+|w_1|)!\ |w_3|!}{n_1^3 (|w_1|+1)^2 (|w_3|+1)^2}\\
     &\le  \  K^{n_1+v_1-2+|w_1|+|w_3|}\   n_1\
     2^{2n_1}\ \left( {n_1 \atop v_1}\right)     \, 
     \frac{(n_1-2+|w_1|)!\ |w_3|!}{n_1^3 (|w_1|+1)^2 (|w_3|+1)^2}\ .
\end{split}
\label{qu12}
\een      
     We  used
      Lemma \ref{decop},  Lemma \ref{compbd}, \eqref{cor1},
      the bound
\[
|\pa^w  \dot C^1(q)| \,\le\,  \,2^{|w|}\, (|w|+1)!
\,\le\,\frac{K^{|w|} \ |w|!}{(|w|+1)^2}\ \ \mbox{ valid for } \ K \ge 16
\ , 
\mbox{ see Lemma \ref{propmu}, \eqref{1-10}, \eqref{3-10} }, 
\]
\[
\mbox{and the inequality }\
\left( {n_1 \atop v'}\right)\left( {n_1 \atop v_1-v'}\right)
\,\le\, 2^{2n_1}\ \left( {n_1 \atop v_1}\right) \ 
\mbox{ for nonnegative integers }  v'\,\le \,v_1 \,\le n_1 \ .
\]
Using \eqref{qu12} we now bound inductively
\eqref{quad}, using Lemma \ref{compbd}, \eqref{cor1}.
We bound the contribution with $n_1 \le n_2\,$ so that the bound we obtain
has to be multiplied by 2. We write for shortness
\[
\sum_{n_i} \,=\,  \sum_{n_{1}+n_{2}=n+2\atop n_1 \le n_2, n_i \in 2 \mathbb{N}}
\ ,\quad
\sum_{w_i} \,=\,\sum_{w_{1}+w_2+w_3=w}\ \ , \quad
\sum_{v_i} \,=\,\sum_{v_1+v_2=v} \ \,.
\]
Using our inductive assumptions and \eqref{qu12},
\eqref{quad} is bounded by
\ben
\begin{split}
&
\frac{2}{n+1}\ 
\sum_{n_{i}}\
  \sum_{w_{i}}\ \sum_{v_i}
c_{w_1,w_2,w_3}\ K^{n_1+v_1-2+|w_1|+|w_3|}\ 
   n_1\, 
     2^{2n_1}\ \left( {n_1 \atop v_1}\right)\,
       \frac{(n_1-2+|w_1|)!\ |w_3|!}{n_1^3 (|w_1|+1)^2 (|w_3|+1)^2}\\
       & \times\ K^{n_2+v_2-2+|w_2|}\,\left( {n_2 \atop v_2}\right)\ 
\frac{(n_2-2+|w_2|)!}{n_2^3 (|w_2|+1)^2}\\
&
\le
\frac{2\,}{n+1}K^{n+v+|w|-2}\, \sum_{n_{i}}\,
n_1  2^{2n_1}
\sum_{w_{i}} 
c_{w_1,w_2,w_3}\ 
 \frac{(n_1-2+|w_1|)!\ |w_3|!}{n_1^3 (|w_1|+1)^2 (|w_3|+1)^2}\
 \frac{(n_2-2+|w_2|)!}{n_2^3 (|w_2|+1)^2}
\sum_{v_i} \left( {n_1 \atop v_1}\right)\!\!
 \left( {n_2 \atop v_2}\right)\\
 &
\le
\frac{2\,}{n+1}K^{n+v+|w|-2}\, \sum_{n_{i}}\
n_1\  2^{2n_1}\ \frac{1}{n_1^3\ n_2^3} \ \left( {n+2 \atop v}\right)
\\
 &
\times\ \sum_{w_{i}} 
\left({ |w|\,\atop  |w_1|+|w_2| }\right)\
\left({ |w_1|+|w_2| \atop  |w_1|}\right)\ 
\frac{|w_3|! \ (n-2+|w_1|)!(n-2+|w_2|)!}
     {(|w_1|+1)^2 (|w_2|+1)^2 (|w_3|+1)^2} \\
 &
\le
\frac{2\,}{n+1}K^{n+v+|w|-2}\, \sum_{n_{i}}\
n_1\  2^{2n_1}\ \frac{1}{n_1^3\ n_2^3} \ \left( {n+2 \atop v}\right)
\frac{(n_1-2)!\ (n_2-2)!}{(n-2)!}\\
&
\times\ \sum_{w_{i}} 
\left({ |w|\,\atop  |w_1|+|w_2| }\right)\
\frac{|w_3|! \ (n-2+|w_1|+|w_2|)!}
     {(|w_1|+1)^2 (|w_2|+1)^2 (|w_3|+1)^2}\\
&
\le
\frac{2\,}{n+1}K^{n+v+|w|-2}\,\sum_{n_{i}}\
n_1\  2^{2n_1}\ \frac{1}{n_1^3\,n_2^3}\ \left( {n+2 \atop v}\right)\
 2^{-(2n_1-2)} \, e \sqrt{2n_1-1}
\\
 &
\times\ \sum_{w_{i}} 
\left({ |w|\,\atop  |w_1|+|w_2| }\right)\ 
\frac{|w_3|!   \ (n-2+|w_1|+|w_2|)!}{(|w_1|+1)^2 (|w_2|+1)^2 (|w_3|+1)^2}
\\
 &
\le
\frac{2^{3.5}\,e }{n+1}\ K^{n+v+|w|-2} \ \left( {n+2 \atop v}\right)\
(n-2+|w|)!\ 
\sum_{n_{i}}\
\frac{1}{n_1^{3/2}\,n_2^3} \sum_{w_{i}} 
 \frac{1}{(|w_1|+1)^2 (|w_2|+1)^2 (|w_3|+1)^2}\
\\
 &
\le
\frac{2^{14.5}\,e }{n+1}\ K^{n+v+|w|-2} \ \left( {n+2 \atop v}\right)\
\frac{1}{n^3}\ \frac{(n-2+|w|)!}{(|w|+1)^2}
\\
 &
\le
2^{16.5}\,e\  K^{-2}\ \frac{1}{ n^3 }
\ K^{n+v+|w|} \ \left( {n+2 \atop v}\right)\
\frac{(n+|w|)!}{(n+2)^3\, (|w|+1)^2}
\ .
\end{split}
\label{qu123}
\een
Here we used in the second estimate
\[ 
\left(a \atop c\right) \left(b \atop d\right)\, \le\
\left(a+b \atop c+d\right)\ \mbox{ for nonnegative integers }
\, a,\ b,\ c,\ d\,; \ a\ge c,\  b\ge d\ ,
\]
implying
\[
c_{w_1,w_2,w_3}\,\le\, \left({ |w|\,\atop  |w_1|+|w_2| }\right)
\left({  |w_1|+|w_2|\,\atop |w_1|}\right)
\]
and the Vandermonde convolution 
\[
\sum_{v_1+v_2=v}  \left( {n_1 \atop v_1}\right)\,\left( {n_2 \atop v_2}\right) 
\,=\,\left( {n +2 \atop v}\right)\ .
\]
In the third estimate we used  the inequality 
\[
(n_1-2 +|w_1|)!\ (n_2-2+|w_2|)!\ {|w_1|+|w_2|\,\choose |w_1|}\
\le\ (n-2+|w_1|+|w_2|)!\ \frac{(n_1-2)!\ (n_2-2)!}{(n-2)!}\ .
\]
In the fourth estimate we used  Lemma \ref{abl}. In the fifth
we applied
\[
(n-2+|w_1|+|w_2|)! 
\left({|w|\,\atop  |w_1|+|w_2| }\right) \ |w_3| ! \le
(n-2+|w|)! \ .
\]
In the sixth we used 
\[
\sum_{n_1+n_{2}=n+2 \atop n_i \in 2 \mn,\, n_2 \ge n_1 \ge 2 } \,
\frac{1}{n_1^{3/2}\, n_2^3} \,\le\, \frac{8}{(n+2)^3} \ , \ 
\sum_{|w_1|+|w_{2}|+|w_3| =|w|} 
\frac{1}{(|w_1|+1)^2(|w_2|+1)^2(|w_3|+1)^2\,}\,\le\, \frac{2^8}{(|w|+1)^2\,}
\]
and in the last one 
\[
\frac{1}{n(n+1)(n-1)} \,\le\,  \frac{4}{(n+2)^3}\ .
\]

\vspace{.3cm}
\noindent
The last but one term in \eqref{cagfew} is inductively
bounded by 
\[
\left|\frac{n-4}{n(n+1)}\,
[\pa^w  \ell_n]_{v,n-v}(\mu; q_{1},\ldots,q_{n})\right|
\, \le\, 
\frac{|n-4|}{n(n+1)}\, \left( {n \atop v} \right)\,
K^{n+v+|w|-2}\,\frac{(n+|w|-2)!}{n^3(|w|+1)^2} 
\]
\[
\,\le\,
\frac{2^4}{3}\,K^{-2} \,\frac{1}{n^2}\, \left( {n \atop v} \right)\,
K^{n+v+|w|}\,\frac{(n+|w|)!}{(n+2)^3(|w|+1)^2}\ \, .
\]
The last  term consists of $n+1$ terms after expanding $d/d\mu$,
each of which  contains $|w|+1$ derivatives. So the inductive bound
gives
\[
(n+1)\ \frac{2}{n(n+1)}\ \left( {n \atop v} \right)\
K^{n+|w|+1-2}\ \frac{(n+|w|+1-2)!}{n^3(|w|+2)^2}
\]
\[
\,\le\,
\frac{8}{n^2}\, K^{-1}\,  \left( {n \atop v} \right)\,
K^{n+v+|w|}\,\frac{(n+|w|)!}{(n+2)^3\,(|w|+1)^2}\ \, .\ \
\]
Assuming that
\[
2\cdot 2\cdot  2^{16.5}\,e\, K^{-2}\, \,+\,
\frac{2^5}{3}\,
K^{-2}\, +\,2^4\ K^{-1}\,\le\, 1 
\]
we find that the r.h.s. of the FE is bounded by
\[
\frac12\, \left[\frac{1}{n}\  \left( {n+2 \atop v} \right)\,
\,+\,
 \left( {n \atop v} \right)\,\right]\frac{1}{n^2}\, 
K^{n+v+|w|}\,\frac{(n+|w|)!}{(n+2)^3\,(|w|+1)^2}\  .
\]
On realises that any $\,K\ge 1100 \,$ is admissible.
Lower values of $\,K\,$ can be achieved when
treating some cases concerning  small values of  $n$ and $|w|$
separately.

\hfill \proofSymbol


\section{Global solutions of the flow equations}
\label{smgs}

According to Proposition \ref{urbild}, \eqref{phipsi},  \eqref{urbildgl},
Proposition  \ref{urbild2}, \eqref{phipsi1},  \eqref{urbildgl2},
 and Proposition \ref{bdphinu}, \eqref{urbildv}
 we can  construct by minimal extension
 - see Definition \ref{minext} - functions 
 $\,\ell_{n+2}(\mu;q_{1},\ldots,q_{n}, q_{n+1}, q_{n+2})\,$
 together with their orthogonal projections onto the spaces
 ${\cal S}_{v,n-v}\,$ which solve the FEs \eqref{cagfe}.
To bound these functions we shall use   \eqref{!!}.
By induction the r.h.s of \eqref{cagfe} is in
$\,{\cal S}^\infty (\mk^{4(n-1)})\,$. 
We call its minimal lifting $\,\psi_{n} \,$,
see Lemma \ref{minlift}, \eqref{lifts}.
The $\,(v,n+2-v)$-component of the  
preimage under $\,\cal C\,$ of $\,\psi_n$, constructed by minimal
extension,  is then denoted as before
by $\,[\Phi_{n+2}(\psi_n)]_{v,n+2-v}\,$,
see \eqref{urbildcomp}. We obtain

{\prop
  Under the assumptions of Proposition \ref{prop4}, 
  the functions $\,[\pa^{w}\Phi_{n+2}(\psi_n)]_{v,n+2-v}\,$
  obey the bounds
\ben
||\,[\pa^{w}\Phi_{n+2}(\psi_n)]_{v,n+2-v}\,||_\infty
\,\le\, 
K^{n+|w|}\ \left( { n+2 \atop v}\right)\ 
\frac{(n+|w|)!}{(n+2)^3\,(|w|+1)^2\,}\ .
\label{bdrhs}
\een

\label{prop5}
}

\vspace{.2cm}

\noindent
\textit{Proof}.
We get from \eqref{!!}, Proposition \ref{bdphinu}, from \eqref{infty},
and from Proposition \ref{prop4}, \eqref{rhsl}
\[
||\,[\pa^w\Phi_{n+2}(\psi_n)]_{v,n+2-v}\,||_\infty
\,\le\, \frac{(n-v+1)}{2}\ K^{n+|w|}\,\frac{(n+|w|)!}{(n+2)^3\, (|w|+1)^2\,}
\]
\[
\times \ \frac12\
\left[\sup_{v \le \la \le n}\, 
\Biggl(\frac{\left( { n+2 \atop v}\right)}{\left( { n \atop \la} \right)} \
\frac{1}{n^3}\ \left( {n+2 \atop \la} \right)\Biggr)\,+\,
\sup_{v \le \la \le n}\, 
\Biggl(\frac{\left( { n+2 \atop v}\right)}{\left( { n \atop \la} \right)} \
\frac{1}{n^2}\ \left( {n \atop \la} \right)\ \Biggr)\right]\ .
\]
Since
\[
(n\!-\!v+1) 
\sup_{v \le \la \le n} 
\Biggl(\frac{\left( { n+2 \atop \la}\right)}
{\left( { n \atop \la} \right)} 
\Biggr) 
\le\,
\sup_{v \le \la \le n}\Biggl( \frac{(n\!-\!v+1)(n+2)(n+1)}
    {(n\!-\!\la+2)(n\!-\!\la+1)}
\Biggr)\le\,
\frac12 (n+2)(n+1)^2 \,\le\, 2\,n^3
\]
and
\[
(n\!-\!v+1) \,\le\, 2\,n^2
\]
we obtain \eqref{bdrhs}.

\hfill \proofSymbol

Our main result is the following\,:

{\thm 
Let be given a two-point function
$\,  \ell_{2}(\mu;q,-q)\, $ which is
uniformly locally analytic (see Remark \ref{analyt})
w.r.t. $\,\mu \in [0, \mu_{\textrm max}]\,$
and  $\,q\,\,$,  and invariant under the Euclidean
group. We denote (by slight abuse of notation)  its minimal lifting,
(see  Definition \ref{dminlift}) as
$\,  \ell_{2}(\mu;q_1,q_2)\,$. It is by construction  symmetric
w.r.t.  $q_1,\,q_2\,$,
and it is uniformly locally analytic w.r.t.
$\,\mu \in [0, \mu_{\textrm max}]\,$ and
w.r.t. $q_1,\,q_2\,$, and invariant under the Euclidean
group if $\,  \ell_{2}(\mu;q,-q)\, $ is so w.r.t. $\mu\, , \ q\,$.

\vspace{.2cm}
\noindent
Assume that there exists $\,K> 0\,$ 
such that  the following bounds hold\,:
\ben
\Bigl| \Bigl|\, [\pa^{w} \,\ell_2]_{v,2-v}(\mu;\cdot) \Bigr| \Bigr| \,\le\, 
\frac{K^{|w|}\ |w| !}{ (|w|+1)^2\,} \quad \forall\ 0 \le v \le 2 \ \mbox{ and }
\ \ w \in \mathbb{N}^3_0\ .
\label{bdthm}
\een
Then there exist solutions
$\,\ell_n(\mu; q_{1},\ldots,q_{n})\,$, $\, n \ge 4\,$,  of the FEs,
constructed inductively from the two-point function
 by minimal lifting and 
minimal extension (see  
Definition \ref{dminlift} and Definition \ref{minext}) 
which  are uniformly locally analytic w.r.t.
$\,\mu\,$ and  w.r.t. $q_1,\,\ldots\, q_n\,$,
invariant under the Euclidean group, and completely symmetric 
under permutations of  $q_1,\,\ldots\, q_n\,$.
Analyticity is now also uniform w.r.t. $\,n \,$.
For $\,n \ge 4\,$ they satisfy the bounds
\ben
\Bigl| \Bigl|
\, [\pa^{w} \ell_n]_{v,n-v}(\mu; \cdot)\,
\Bigr| \Bigr| \,\le\, 
K^{n-2+|w|}\ \left( {n \atop v} \right)\
\frac{(n-2+|w|)!}{n^3\, (|w|+1)^2\,}\ , \ \ 
 0 \le v \le n\,, \ \ w \in \mathbb{N}^{n+1}\ .
\label{erwbd}
\een
\label{thm1}
}

\vspace{.2cm}

\noindent
\textit{Proof}.
Uniform local analyticity w.r.t. $\,\mu\,$ and w.r.t. 
to the momenta $q_i$ follow inductively from the 
FEs since analytic functions form a ring, since $\dot C^{1}(q)\,$
is analytic w.r.t. $\mu\,$ and $q\,$, and since minimal lifting
and minimal extension
obviously respect analyticity. Similar arguments assure
symmetry and invariance under the Euclidean group.
Minimal lifting and minimal extension
are also taylored to preserve Bose symmetry and Euclidean invariance.
The bounds \eqref{erwbd} follow inductively 
from Proposition \ref{prop4} and Proposition \ref{prop5}.  
\hfill \proofSymbol

\vspace{.2cm}

{\prop (Uniqueness)
  \label{uniqueness}
  Let $\,\ti \ell_n (\mu; q_{1},\ldots,q_{n})\,$ be a system 
  of solutions of the FEs uniformly locally analytic w.r.t.
$\,\mu\,$ and  w.r.t. $q_1,\,\ldots\, q_n\,$,
invariant under the Euclidean group, and completely symmetric 
under permutations of  $q_1,\,\ldots\, q_n\,$. If 
the system  $\,\ti \ell_n (\mu; q_{1},\ldots,q_{n})\,$ satisfies
the same boundary conditions as the system
$\,\ell_n (\mu; q_{1},\ldots,q_{n})\,$ from Theorem \ref{thm1}, 
  \ben
  \ti \ell_n (0; q_{1},\ldots,q_{n}) = \ell_n(0; q_{1},\ldots,q_{n})
  \quad \forall n,\ q_i\ , 
  \een
  then
  \ben
  \ti \ell_n (\mu; q_{1},\ldots,q_{n}) = \ell_n(\mu; q_{1},\ldots,q_{n})
\quad  \forall \,\mu \in [0, \mu_{\textrm max}]\, ,\ q_i \in \mathbb{\mr}^4\ .
  \een
 }

\noindent
\textit{Proof}. 
We consider the system
\[
\De \ell_n (\mu; q_{1},\ldots,q_{n}) =
\ell_n (\mu; q_{1},\ldots,q_{n}) \,-\, \ti \ell_n (\mu; q_{1},\ldots,q_{n}) \ .
\]
These functions have the same analyticity properties
as the $\ell_n$ and $\ti\ell_n$.   
By telescoping the difference of the
FEs for $\,\pa_\mu \ell_n\,$ and  $\,\pa_\mu \ti \ell_n\,$,
one realises that the system $\,\De \ell_n\,$ satisfies
a system of \textit{linear} FEs depending on 
the  $\,\ell_n\,$ and  $\,\ti \ell_n\,$. The boundary conditions
for this system at $\,\mu=0\,$ vanish identically.
By integrating the  functions $\pa_\mu \De \ell_n$ in a neighbourhood
of $\,\mu=0\,$ one finds inductively that they vanish
to infinite order at $\mu=0$. So by local analyticity they vanish
identically
in a neighbourhood of  $\,\mu=0\,$. By a compactness argument they then
vanish for all $\,0 \,\le\, \mu \,\le\, \mu_{\textrm{max}}\,$.
\hfill \proofSymbol

\vspace{.2cm}
\noindent
 For ultraviolet asymptotically free solutions 
 the effective couplings, as  functions
of the energy scale, tend to zero if the energy scale
tends to infinity, see ~\cite{Kopper} for more details.
The energy scale is given by $\,\al^{-1/2}\,$.
The bare couplings are fixed at the energy scale 
$\,\alo^{-1/2}\,$ corresponding to the value 
$\,\mu =0\,$. 
In the ultraviolet limit $\,\alo \,\to\, 0\,$, 
the physcial scale $\,\al=1\,$ 
corresponds to $\,\mu_{\textrm{max}} \,\to\, \infty\,$.
We  now assert that there exist asymptotically free,
gobally bounded solutions of \eqref{CAGFEsexpand}
or  \eqref{cagfe} which satisfy the assumptions
of Theorem \ref{thm1}.

{\cor 
\label{thlimits} (Example of an UV asymptotically free solution).
We consider  the two-point function 
\ben
 {\ell}_2(\mu;q,-q)= -\de(\mu)\, ,\   \de(\mu_{\max})=\,\de\, , \  
\pa_\mu  \de(\mu)= \be \de^2(\mu)\, , 
 \ 0 < \de,\, \be \, ,
  \  \mu \in [0,\,\mu_{\textrm max}]\ .
\label{f2asy1}
\een
Then 
\ben
\ell_2(\mu_{\textrm{max}};q_1,q_2)
\,=\, -\,\de\ ,\quad
\pa_\mu^l\, \ell_2(\mu_{\textrm{max}};q_1,q_2)\,=\, -\,\be^l\, l!\ \de^{l+1}\ \ ,
\label{limits}
\een 
and we have
\ben
\mbox{UV asymptotic freedom }: \quad \lim_{\mu_{\textrm{max}}\to \infty} \pa_\mu^l\,
\ell_n(0;q_1,\ldots,q_n)\,=\, 0\quad \forall\ l \ge 0,\,\  n \ge 2\, \ .
\label{nlimits}
\een 
}

\vspace{.2cm}

\noindent
\textit{Proof}.
The well-known solution of $\;\pa_\mu  \de(\mu)= \be \de^2(\mu)\,$, 
satisfying  
$\,\de(\mu_{\textrm{max}})
\,=\, \,\de\,$
is
\ben
\de(\mu)\,=\, \frac{\de}{1\,+\,\de\,  \be\,(\mu_{\max}-{\mu})}\ \ .  
\label{f2asy2}
\een
Since $\, \pa_\mu^l\,\de(\mu)
\,=\, \,\be^l\, l!\ \de^{l+1}(\mu)\, $, the
two-point function \eqref{f2asy1}  satisfies the
assumptions of Theorem \ref{thm1}.
Furthermore, because
\ben
\Bigl|\Bigl|\, [\pa^w \ell_2]_{v,2-v}(\mu;\cdot)\, 
\Bigl|\Bigl|_\infty\, \le \, \de^{1+|w|}(\mu)\
K^{w+v}\,\frac{|w|!}{2^3 (|w|+1)^2\,}\ , \quad 0 \le v \le 2\ ,
\label{asy2ptbd}
\een
uniformly in the momenta, 
it is immediate to verify inductively that the bounds 
on the r.h.s. of the FEs \eqref{cagfew}
from Proposition \ref{prop4}, 
\eqref{nptbd} and \eqref{rhsl},
also hold  on multiplication by 
a factor of $\,\de^2(\mu)\,$.
Note that we do not get a factor of $\,\de^{2+|w|}(\mu)\,$  
because the  derivatives $\,\pa^w\,$ may apply to
to the factor $\,\dot C^1(q)\,$ on the r.h.s. of the FEs
in which case they do not produce additional decrease w.r.t.
$\,\de(\mu)\,$.
From this we conclude by the usual inductive argument  that
\[
\Bigl|\Bigl|\, \,\pa^w\,\ell_n(\mu;\cdot)
\,\Bigl|\Bigl|_\infty\,\le\,{\cal O}(\de^2(\mu))
\  \Rightarrow\
\lim_{\mu_{\textrm{max}}\to \infty} \pa^w\,
\ell_n(0;q_1,\ldots,q_n)\,=\, 0\quad \forall\ |w| \ge 0,\,\  n \ge 2\, \ .
\]
\hfill \proofSymbol

Undoing the scalings \eqref{scaling1}, \eqref{scaling2} we find for the
connected amputated Schwinger functions the same scaling behaviour
\ben
{\L}^{\al_0,1}_{n}(p_{1},\ldots,p_{n})\,=\,
{\cal O}(\mu_{\textrm{max}}^{-2}) \ ,\quad
\mu_{\textrm{max}}\,=\, \ln \frac{1}{\al_0}\ , 
\label{backscaling}
\een
remembering that for $\,\mu=\mu_{\textrm{max}}\,$ we have $\,\alpha =1\,$.

\vspace{.3cm}

The sign of $\,{\ell}_2(0;q,-q)\,$ in \eqref{f2asy1}
is in agreement with the sign of the mass counter term in 
perturbative $\vp^4$-theory  to lowest order.
The function $\,\beta\, \de^2(\mu)\,$ plays a role similar to that of 
the lowest order  $\beta$-function in perturbative quantum field theory,
where the contributions are ordered according to the number of loops.
But in our example the four-point function
decreases proportionally to the \textit{square} of the logarithmic energy
scale, whereas this decrease is linear for standard lowest order
perturbative asymptotically free solutions.
If one wants to get closer to the perturbative scheme 
one should start by 
imposing an asymptotically free four-point function with
linear decrease and then calculate the corresponding
two-point function from the FEs. Afterwards one could
start again the inductive construction from there.
Ultimately it will be  decisive which choices (if any)
satisfy the requirements for a full-fledged quantum field
theory, in particular positivity and unitarity.


\section{Appendix}
\label{sec:App}

{\lemma 
Let $\, a, b \in  \mn\,, \ a \le b\,$. Then
\ben
2^{2a}\, \frac{a!\ b!}{(a+b)!} \, \le \, e \sqrt{2a-1} \, \le \,
e \sqrt{a+b-1} 
\ .
\label{ab}
\een   
\label{abl}
} 
\vspace{.2cm}

\noindent
\textit{Proof}.
It is obvious that
$\,\frac{a!\ b!}{(a+b)!}\,$ is decreasing with $\,b\,$ for
 $\,a\,$ fixed. It is therefore sufficient to prove 
the bound for  $\,a\,= b\,$. We have
\[
2^{2a}\, \frac{(a!)^2}{(2a)!} \,=\,  \frac{[(2a)!!]^2}{(2a)!}\, =\,
\prod_{k=1}^{a} \frac{2k}{2k-1}\,=\,\prod_{k=0}^{a-1}(1 +\frac{1}{2k+1})\ , 
\]  
\[
\ln\left(\prod_{k=0}^{a-1}(1 +\frac{1}{2k+1}) \right) \le
\sum_{k=0}^{a-1}\frac{1}{2k+1} \,\le\, 1+ \int_{0}^{a-1} \frac{dx}{2x+1}
\,=\, 1 \,+\,\frac12 \ln(2a-1)
\]
\[
\Rightarrow\ \ \, 2^{2a}\, \frac{a!\ b!}{(a+b)!} \, \le \,
 e \,\sqrt{2a-1}\ . 
\]
 \hfill \proofSymbol

 \vspace{.2cm}
{\lemma 
For $\,0 <  \alo \le \al\,$, 
setting  $\,\mu =\ln(\frac{\al}{\alo})\,$, we have  
\ben
|\,\pa_{\mu}^l \, \dot{C}^{\al}(p)\,|\, \le\,  2^{l}  (l+1)!\ \ ,\quad
|\,\pa_{\mu}^l \, \dot{C}^{\al}(p)\,|\, \le\, 
2\cdot  4^{l}\,l!\,
\,e^{-\frac{\al}{2} p^2}\ \, ,
\label{1-10}
\een
and 
\ben
|\,\pa_{\mu}^l \,[\al^2\,\dot{C}^{\al}(p)]\,|\, \le\, 
\al^2\, 2^{l}\cdot e\cdot (l+1) !
\ \ ,\quad
|\,\pa_{\mu}^l \,[\al^2\, \dot{C}^{\al}(p)]\,|\, \le\, 
\al^2\,2\cdot e^{1/2}\,\cdot 4^{l}\,l!\,
\,e^{-\frac{\al}{2} p^2}\ \ .
\label{2-10}
\een
\ben 
\left|(\frac12 \sum_{\nu=1}^4 p_{\nu} \pa_{p_{\nu}})^l\,\dot{C}^{\al }(p)\right| \le
2^{l}  (l+1)!\ \ ,\quad
\left|(\frac12 \sum_{\nu=1}^4 p_{\nu} \pa_{p_{\nu}})^l\,\dot{C}^{\al }(p)\right|
\le
2\cdot  4^{l}\,l!\,
\,e^{-\frac{\al}{2} p^2}\ \, ,
\label{3-10}
\een
\label{propmu}
}

\noindent
\textit{Proof} . 
For $\,l \ge 1$ we find inductively in  $\,l$
\[
\pa_{\mu}^l \, \dot{C}^{\al}(p)\,=\,
\sum_{\nu =1}^l a_{l,\nu} (-\al p^2)^\nu \,e^{-\al p^2}\ ,
 \]
 where the coefficients $\,a_{l,\nu}\,$, $0 \le  \nu \le l\,$,
 obey the  relations
\[
a_{l,l} \,=\, 1\,, \quad a_{l,0}\,=\,0\,, \quad
a_{l+1,\nu} \,=\, \nu\, a_{l,\nu} \,+\,a_{l,\nu-1}
\ \
\mbox{with}\ \  a_{l,-1} :=0\  .
\]
It then follows by induction in $\,l+\nu\,$ that
\[
|\,a_{l,\nu} \,| \,\le\, 2^l \frac{ l!}{\nu !}
\]
so that
\[
|\,\pa_{\mu}^l \, \dot{C}^{\al}( p)\,| \,\le\,
\sum _{\nu =1}^l 2^l \,\frac{ l!}{\nu !}\, [\al  p^2]^\nu \,e^{-\al p^2}
 \,\le\,
\sum _{\nu =1}^l \frac{2^l  l!}{\nu !}\, \nu ! \,\le \, 2^l \, (l+1)!
\]
and also
\[
|\pa_{\mu}^l \, \dot{C}^{\al }(p)| \,\le\,
\sum _{\nu =1}^{l-1} \frac{2^l l!}{\nu !}\, 
\Bigl[ \al p^2\,\Bigr]^\nu \ e^{- \al p^2} \,+\,\Bigl[ \al p^2\,\Bigr]^l
\ e^{- \al p^2}
\,\le\, \sum _{\nu =1}^{l-1} \frac{ 2^l l!}{\nu !}\ 
2^{\nu}\ \nu!\  e^{-\frac{\al}{2} p^2}\,+\, 2^l\ l!\  e^{-\frac{\al}{2} p^2}
\]
\[
\le\,
\Bigl(\sum _{\nu =1}^{l-1} 2^{\nu}\,+\,1\Bigr)\,\,2^l\ l!\  e^{-\frac{\al}{2} p^2}
\, \le\,
4^{l}\,l !\
e^{-\frac{\al}{2} p^2}\  .
\]
These bounds do not take into account the alternating signs in the sum.\\ 
The bounds \eqref{2-10} then follow from 
\[
\pa_{\mu}^l \,\al^2 \,=\, 2^l \al^2\ ,\quad
\sum_{\la \le l} \left( l \atop \la\right)\, 2^\la \,2^{l-\la} (l-\la+1)!
\ \le\  2^l\  l! \,\sum_{\la \le l}
\frac{l-\la +1}{\la !} \le \,2^l (l+1)! \ e \ \,,
\]
\[
\sum_{\la \le l} \left( l \atop \la\right)\,  2^\la  \ 4^{l-\la} (l-\la)!
\ \le\ 
 4^l\ l! \,\sum_{\la \le l} \frac{2^{-\la}}{\la !}\, \le \, 4^l\ l!\  e^{1/2} \ \, .
 \]
Finally  \eqref{3-10} follows from 
 \[
 (\frac12 \sum_{\nu=1}^4
 p_{\nu} \pa_{p_{\nu}})^l \dot{C}^{\al }(p) = \pa_\mu^l \dot{C}^{\al }(p)\ .
 \]
 \hfill  \proofSymbol


\vspace{.3cm}

\noindent
 \textbf{Data Availability}
 Data sharing is not applicable to this article as no
 datasets were generated or analysed during
the current study.

\vspace{.2cm}

\noindent
\textbf{Declarations}

\vspace{.2cm}

\noindent
\textbf{Conflict of interest}
The author has no conflict of interest to
declare that is relevant to the content of this
article.


\vspace{.3cm}

\addcontentsline{toc}{chapter}{Bibliography}
\begin{thebibliography}{10}

\bibitem{Aizenman82}
M. Aizenman, 
``{Geometric analysis of $\phi^4$ fields and Ising models, 
Parts I and II}'',  {\em Commun.Math.Phys.} {\textbf 86} (1982) 1-48.  
%
\bibitem{ADC2021} M. Aizenman and H. Duminil-Copin,
``{Marginal triviality of the scaling limits of critical
4D Ising and $\phi_4^4$ models}'', 
 {\em Ann.Math.}, {\textbf 194} (2021), 163-235. 
%
\bibitem{Brydges} D. Brydges, ``{A short course on cluster expansions}'',
in {\em \, Les Houches Summer School 1984\,: Critical phenomena,
random systems, gauge theories}, {\em Elsevier 1986}.
%
\bibitem{BBS} D. Brydges, R. Bauerschmidt and G. Slade,
``{\em Introduction to a Renormalisation Group Method} '', 
{\em Springer Verlag 2019}.
%
\bibitem{BK} D. Brydges, T. Kennedy,
``{Mayer expansions and the Hamilton-Jacobi equation}'',
{\em Journ.Stat.Phys.} {\textbf 48} (1987) 19-49.
%
\bibitem{Du} P. Duch, ``{Construction of
  Gross-Neveu model using Polchinski flow equation}'',
  preprint arXiv 2403.1852.
%
\bibitem{EGK} A.N. Efremov, R. Guida, Ch. Kopper, ``{Renormalization
of $SU(2)$ Yang-Mills theory with flow equations}'', {\em Journ.Math.Phys.}
{\textbf 58} (2017) 093503.     
%
\bibitem{Felder}
G. Felder, ``{Renormalisation group in the local potential
approximation}'', {\em Commun. Math.Phys.} {\textbf 111} (1987)
  101-121.
%
\bibitem{FroebH} J. Fr\"ob, J. Holland, S. Hollands, 
``{All-order bounds for correlation functions of gauge-invariant 
operators in Yang-Mills theory}'', {\em Journ.Math.Phys.}
{\textbf 57} (2016) 122301.   
%
\bibitem{Froehlich82} 
J. Fr\"ohlich, ``{On the triviality of $\la \phi^4_4$ theories 
and the approach to the critical point in $d \ge 4$ dimensions}'',
{\em Nucl.Phys} {\textbf B 200 [FS4]} (1982) 281-296.
%
%
\bibitem{FFS} R.~Fern\'andez , J.~ Fr\"ohlich , A.D.~Sokal
  ``{\em Random Walks, Critical Phenomena, and Triviality
  in Quantum Field Theory}, {\em Springer Verlag 1992}
  %
\bibitem{GK} K. Gawedzki, A. Kupiainen, 
``{Non-Gaussian fixed points of the block spin transformation.
Hierarchical model approximation}'',
{\em Commun.Math.Phys.} {\textbf 89} (1983) 191-200.   
\bibitem{Glimm}J. Glimm and A. Jaffe, ``{Quantum Physics: 
A Functional Integral Point of View}'',  {\em Springer Verlag 1987}.
%
\bibitem{Gr} R.L. Greenblatt, ``{Constructing a weakly-interacting fixed point
 of the fermionic Polchinski equation}'', {\em Journ.Math.Phys.} 
{\textbf 65} (2024) 112301.
%
\bibitem{Gould}J. Quaintance, Combinatorial Identities: Table I: 
Intermediate Techniques for Summing Finite Series,
From the seven unpublished manuscripts of H. W. Gould, 2010
http://www.math.wvu.edu/~hgould/Vol.4.PDF.
\bibitem{HH} K.~Halpern, K.~Huang,
``{Nontrivial directions for scalar fields}''
{\em Phys.Rev.} {\textbf D 53} (1996) 3252-3259.
%
\bibitem{Hurd:1989} T.R.~Hurd, 
``{A renormalization group proof of perturbative renormalizability}'',
{\em Commun.Math.Phys.} {\textbf 124} (1989)
  153-168.
%
\bibitem{Keller:1991bz}
  G.~Keller and Ch.~Kopper, ``{Perturbative renormalization
    of composite operators
  via flow equations. 1.}'', {\em Commun.Math.Phys.} {\textbf 148} (1992)
  445-468.
%
\bibitem{Keller:1990ej}
G.~Keller, Ch.~Kopper, and M.~Salmhofer, ``{Perturbative renormalization and
  effective Lagrangians in $\Phi^4$ in four-dimensions}'', {\em Helv.Phys.Acta}
  {\textbf 65} (1992)  32-52.
\bibitem{Kopper}
Ch. Kopper, 
``{Asymptotically free solutions of the scalar mean field
  flow equations}'',  {\em Ann. Henri Poincar\'e} {\textbf 23} (2022)
3453-3492.  
\bibitem{KW}
  Ch. Kopper, P. Wang, ``{Triviality proof for mean field 
  $\vp^4$-theories in four dimensions}'',
  {\em J.Math.Phys.} {\textbf 66}  (2025), 062302 (1-35).
%
\bibitem{MayerMontroll}
J.E. Mayer, E. Montroll, ``{Molecular distributions}'', 
{\em J.Chem.Phys.} {\textbf 9} (1941) 2-16.
%
\bibitem{Morris} I. Hamzaan Bridle, Tim R. Morris
``{The fate of non-polynomial interactions in
scalar field theory}'', {\em Phys.Rev.} {\textbf D 94} (2016) 065040.  
%
\bibitem{Muller:2002}
V.~F. M{\"u}ller, ``{Perturbative renormalization by flow equations}'', {\em
  Rev.Math.Phys.} {\textbf 15} (2003)  491-558.
%
\bibitem{OSa} P.~Obernolte, M.~Salmhofer,
  ``{\em Integrating Polchinski's equation by
  convergent binary tree expansions}'', arXiv:2606.21644  
%
\bibitem{OS}
    K.~Osterwalder, R.~Schrader, ``{\em Axioms for Euclidean Green's
    Functions II}'', {\em Commun.Math.Phys.} {\textbf 42} (1975) 281-305.
\bibitem{Polchinski84}
J.~Polchinski, ``{Renormalization and Effective Lagrangians}'', {\em
  Nucl.Phys.} {\textbf B231} (1984)  269-295.
%
\bibitem{Riv:PCR} V. Rivasseau, ``{\em From Perturbative to Constructive
Renormalization}'',  {\em Princeton University Press 1991}.
%
\bibitem{Simon} B. Simon,  
``{\em $P(\phi_2)$ Euclidean Quantum Field Theory}'',
{\em Princeton University Press 1974}.
%
\bibitem{Wegner72}
F.J. Wegner and A.~Houghton, ``{Renormalization group equation for critical
  phenomena}'', {\em Phys.Rev.} {\textbf  A8} (1973)  401-412.
%
\bibitem{Wilson71}
K.G. Wilson, ``{Renormalization group and critical phenomena. 1.
  Renormalization group and the Kadanoff scaling picture}'', {\em Phys.Rev.}
  {\textbf B4} (1971)  3174-3183.\\
K.G. Wilson, ``{Renormalization group and critical phenomena. 2. Phase space
  cell analysis of critical behavior}'', {\em Phys.Rev.} {\textbf{ B4}} (1971)
  3184-3205.
%
\bibitem{WF72}
K.G. Wilson, M.E. Fisher,
 ``{Critical exponents in 3.99 dimensions}'', {\em Phys.Rev.Lett}
  {\textbf{28}} (1972)  240-243.




\end{thebibliography}
\end{document}